\documentclass[11pt]{revtex4}
\usepackage{mathtools}
\usepackage{hyperref}
\usepackage{pdflscape}
 \usepackage{graphics}
 \usepackage{subfigure}
\usepackage{epsfig}
 \usepackage{amsmath,amsfonts}
 \usepackage{amssymb}
 \usepackage{color}
 \usepackage{ulem}
\begin{document}
\title{Polarization modes of gravitational waves in Palatini-Horndeski theory}

\author{Yu-Qi Dong$^{a,b}$}
\email{dongyq21@lzu.edu.cn}

\author{Yu-Xiao Liu$^{a,b}$} % \footnote{Corresponding author}
\email{liuyx@lzu.edu.cn, corresponding author}

\affiliation{
$^{a}$Lanzhou Center for Theoretical Physics,
	Key Laboratory of Theoretical Physics of Gansu Province,
	School of Physical Science and Technology,
    Lanzhou University, Lanzhou 730000, China \\
$^{b}$Institute of Theoretical Physics \& Research Center of Gravitation,
    Lanzhou University, Lanzhou 730000, China
}

\begin{abstract}
\textbf{Abstract:} In this paper, the polarization modes of gravitational waves in Horndeski gravity are studied under the Palatini formalism. After obtaining the linearized equation of perturbations in Minkowski background, we find that the polarization modes of gravitational waves depend on the selection of the theoretical parameters. The polarization modes can be divided into quite rich cases by parameters. In all cases of parameter selection, there are $+$ and $\times$ modes propagating at the speed of light but no vector modes. The only difference from general relativity is scalar modes, especially the scalar degrees of freedom can be 0, 1 or 2 in different cases. The appropriate parameter cases can be expected to be selected in the detection of gravitational wave polarization modes by Lisa, Taiji and TianQin in the future.
\end{abstract}
	
\maketitle

\section{Introduction}
\label{sec: intro}
It is well known that LIGO and Virgo have {detected} {gravitational waves \cite{Abbott1,Abbott2,Abbott3,Abbott4,Abbott5}, which implies the arrival of the gravitational wave astronomy era.} Since then, in addition to receiving electromagnetic radiation from space, we have another important means to understand the universe by detecting gravitational waves. This breakthrough enables us to test various candidate modified gravitational theories by observing the polarization modes of gravitational waves. %\textcolor{blue}{Especially, the theoretical methods }\textcolor{red}{of} \textcolor{blue}{ detecting polarization modes of gravitational waves were investigated in  Refs. \cite{Eardley,Atsushi Nishizawa,Kazuhiro Hayama,Yuki Hagihara,H. Takeda,Peter T. H. Pang,K. Chatziioannou,Maximiliano Isi1,Maximiliano Isi2,Hiroki Takeda} and observational results of detecting polarization modes were given in Refs. \cite{B. P . Abbottet al.1,B. P . Abbottet al.2,B. P . Abbottet al.3,B. P . Abbottet al.4}.}
%\textcolor{blue}{Refs. \cite{Eardley,Atsushi Nishizawa,Kazuhiro Hayama,Yuki Hagihara,H. Takeda,Peter T. H. Pang,K. Chatziioannou,Maximiliano Isi1,Maximiliano Isi2} studied the theoretical methods of detecting polarization modes of gravitational waves and Refs. \cite{B. P . Abbottet al.1,B. P . Abbottet al.2,B. P . Abbottet al.3,B. P . Abbottet al.4,Hiroki Takeda} are experimental examples of detecting polarization modes.}

After linearly approximating the weak gravitational field, we know the gravitational waves in general relativity has only $+$ and $\times$ polarization modes \cite{MTW}. In 1973, Eardley, Lee and Lightman pointed out that the general four-dimensional metric theory has six possible polarization modes of gravitational waves, and the gravitational wave polarization modes predicted by various specific four-dimensional metric theories are only a subset of these six modes \cite{Eardley}. In recent years, since the three detectors of the LIGO-Virgo collaboration provide the possibility of cross-matching arrival time observations, preliminary work to  {detect polarization modes} has already started {\cite{B. P . Abbottet al.2,B. P . Abbottet al.3,Abbott4,Abbott5,B. P . Abbottet al.4,Hiroki Takeda}.}
{Especially, the theoretical methods of detecting polarization modes of gravitational waves were investigated in Refs. \cite{Eardley,Atsushi Nishizawa,Kazuhiro Hayama,H. Takeda,K. Chatziioannou,Maximiliano Isi1,Maximiliano Isi2} and observational results of detecting polarization modes were given in Refs. \cite{B. P . Abbottet al.1,B. P . Abbottet al.2,B. P . Abbottet al.3,B. P . Abbottet al.4,Yuki Hagihara,P. T. H. Pang,Hiroki Takeda}.}

We have known the polarization modes of gravitational waves predicted by some modified gravities, like Brans-Dicke, $f(R)$, Horndeski, Einstein-ather, TeVeS, Horava, STVG, $f(T)$, dCS and {EdGB theories} \cite{Eardley,f(R,Horndeski,TeVeS,Horava,STVG,fT,dCS and EdGB}. In these studies, the Newman-Penrose formalism was often used in the case of plane gravitational waves propagating at the speed of light \cite{Eardley,N-P}. The polarization modes predicted by various theories are different, so it can be expected that some of them will be excluded from the detection of polarization modes of gravitational waves in Lisa, Taiji and TianQin {\cite{lisa,taiji,Lisa-taiji,tianqin}} in the future.

{Recently, a tentative %strong
 indication for scalar transverse gravitational waves was reported in Ref.~\cite{Z. Chen}.} If it is confirmed in the future, the result will strongly suggest that there should be a scalar degree of freedom in the gravitational theory that describes our world. In the metric formalism, Horndeski theory is the most general scalar-tensor theory that can derive second-order field equations \cite{Horndeski2}. The action of Horndeski theory is \cite{Horndeski3}
\begin{eqnarray}
	\label{action}
	S\; = \; \int d^4x \sqrt{-g}
	\;\Bigl(\mathcal{L}_{2}+\mathcal{L}_{3}+\mathcal{L}_{4}+\mathcal{L}_{5}\Bigl)
	\label{actioneq},
\end{eqnarray}
where,
\begin{eqnarray}
	\label{L2}
	\mathcal{L}_{2}&=&K(\phi,X),
	\\
	\label{L3}
	\mathcal{L}_{3}&=&-G_{3}(\phi,X)\tilde{\Box}\phi,
	\\
	\label{L4}
	\mathcal{L}_{4}&=&G_{4}(\phi,X)\mathnormal{R}+G_{4,X}(\mathnormal{\phi},X)\left[\left({\tilde{\Box}\phi}\right)^{2}-\left(\nabla_{\mu}\nabla_{\nu}\phi\right)\left(\nabla^{\mu}\nabla^{\nu}\phi\right)\right],
    \\
    \label{L5}
    \nonumber
    \mathcal{L}_{5}&=&G_{5}(\phi,X)\left(\mathnormal{R}_{\mu\nu}-\frac{1}{2}\mathnormal{g}_{\mu\nu}\mathnormal{R}\right)\nabla^{\mu}\nabla^{\nu}\phi
    \\ \nonumber
    &-&\frac{1}{6}G_{5,X}(\phi,X)\left[
    \left(\tilde{\Box}\phi\right)^{3}-3\tilde{\Box}\phi
    \left(\nabla_{\mu}\nabla_{\nu}\phi\right)
    \left(\nabla^{\mu}\nabla^{\nu}\phi\right)\right.
    \\
    &+&2\left.\left(\nabla^{\lambda}\nabla_{\rho}\phi\right)\left(\nabla^{\rho}\nabla_{\sigma}\phi\right)\left(\nabla^{\sigma}\nabla_{\lambda}\phi\right)\right].
\end{eqnarray}
Here $\tilde{\Box}=\nabla^{\mu}\nabla_{\mu}$, $X=-\frac{1}{2}\partial_{\mu}\phi\partial^{\mu}\phi$, and $ K, G_{3}, G_{4}$ and $G_{5}$ are functions of the variables $\phi$ and $X$. $G_{4,X}$ and $G_{5,X}$ are respectively the partial derivatives of $G_{4}$ and $G_{5}$ with respect to the variable $X$. The remaining partial derivatives are still represented by this {notation}. For example $K_{,{\phi\phi}}$ represents the second-order partial derivative of $K$ to the variable $\phi$.

We have known the polarization modes of Horndeski theory in the metric formalism  \cite{Horndeski}.  The basic result is: besides $+$ and $\times$ modes, there is one more scalar mode which is the mixture of the breathing and longitudinal modes. And in the massless limit {for the scalar mode}, only the breathing mode survives. {In the metric Horndeski theory, the probability of Landau damping for gravitational scalar waves \cite{Fabio Moretti 1} and the polarization modes of gravitational waves under arbitrary background \cite{Harvey S. Reall} were also studied.}

 Different from general relativity, Horndeski theory in Palatini formalism is different from that in metric one. It can even derive higher-order field equations \cite{Helpin}. {At present, to the best of our knowledge, there are only a few studies on Palatini-Horndeski theory \cite{Helpin,Helpin2}.} It was pointed out that Palatini-Horndeski theory may have interesting
{results. For example,  cosmology of Palatini-Horndeski theory is different from that of metric Horndeski theory and their stability properties are different \cite{Helpin}. It is expected that the Palatini formalism will provide us rich phenomena that does not exist in the metric formalism \cite{Shimada}.
In addition, the gravitational wave event GW170817 together with the gamma ray burst GRB 170817A strictly requires tensor gravitational waves to propagate at the speed of light in the universe with a high accuracy \cite{B. P. Abbott000,B. P. Abbott111}. So, when calculating the deviation of the speed of tensor  gravitational waves from the speed of light in the Friedmann-Robertson-Walker (FRW) metric, it was found that the parameter space of metric Horndeski theory is severely constrained after GW170817. Through this calculation, Ref. \cite{P. Creminelli and F. Vernizzi} showed that the possible subclasses of metric Horndeski theory remain {$S= \int d^4x \sqrt{-g} [K(\phi,X)-G_{3}(\phi,X)\tilde{\Box}\phi+G_{4}(\phi)R]$}. More research on the metric Horndeski theory after GW170817 can be seen in Refs. \cite{C. D. Kreisch,Y. Gong 00}. This constraint limits the application of scalar-tensor theories. Therefore, it is necessary to find scalar-tensor theories beyond the metric Horndeski framework. Bahamonde \textit{et al}. tried to answer this question by establishing teleparallel Horndeski theory \cite{S. Bahamonde 1,S. Bahamonde 2}. The metric Horndeski theory is included in the teleparallel framework as one of many subclasses. Thus, taking Horndeski theory in Palatini formalism may be another idea. Therefore, it is necessary to study the polarization modes of Horndeski theory in the Palatini formalism.
}
 %In addition, the parameters of Horndeski theory are severely constrained after GW170817 \cite{C. D. Kreisch,Y. Gong 00}. Bahamonde et al. showed that Horndeski theory can be recast by using teleparallel formalism \cite{S. Bahamonde 1,S. Bahamonde 2}. Taking the Palatini formalism may be another idea to revive the Horndeski framework. Therefore, it is necessary to study the polarization modes of Horndeski theory in Palatini formalism.

In the Palatini formalism, the polarization modes of gravitational waves have been studied in some gravitational theories, such as Holst-$f(R)$ \cite{F. Bombacigno} and generalized Brans-Dicke theory \cite{J.Lu}. In Ref. \cite{S. Bahamonde}, the polarization modes of gravitational waves in teleparallel Horndeski theory have also been studied. In this paper, we will study the polarization modes of gravitational waves in Palatini-Horndeski theory {in the range of a linear analysis and the possible parameters of Palatini-Horndeski theory after GW170817 will be studied in our future work.} In Minkowski background, we put the metric signature $(-,+,+,+)$ and assume that the space-time has a background solution
\begin{eqnarray}
	\label{background}
	 g_{\mu\nu}=\eta_{\mu\nu},\quad
	 \Gamma^{\lambda}_{\mu\nu}=\mathop{\Gamma}^{0}~\!\!^{\lambda}_{\mu\nu},\quad
	 \phi=\phi_{0},
\end{eqnarray}
where $\eta_{\mu\nu}$ is the Minkowski metric {and the background scalar field $\phi_{0}$ is a constant. The notation ``0" above $\Gamma^{\lambda}_{\mu\nu}$ in Eq. (\ref{background}) represents the background connection, whose components are also constants. Incidentally, for a given background solution, after redefining the scalar field by $\phi\rightarrow\phi+\phi_{0}$, we can always make $\phi_{0}=0$.} %{The notation ``0" above $\Gamma^{\lambda}_{\mu\nu}$ and the subscript ``0" of $\phi$ represent the background connection and the background scalar field, respectively. %They are all coordinate independent constants.} %The background connection and the background scalar field are coordinate independent constants.
We will investigate perturbations of the background solution
\begin{eqnarray}
	\label{perturbation}
	g_{\mu\nu}=\eta_{\mu\nu}+h_{\mu\nu},\quad	\Gamma^{\lambda}_{\mu\nu}=\mathop{\Gamma}^{0}~\!\!^{\lambda}_{\mu\nu}+\Sigma^{\lambda}_{\mu\nu},\quad
	\phi=\phi_{0}+\varphi.
\end{eqnarray}
Our goal is to obtain the linearized field equations of perturbations and then analyze the polarization modes of gravitational waves allowed by these equations. Therefore, in Sec. \ref{sec: 2}, we write the first-order terms of perturbations in the gravitational action and obtain the equations satisfied by the background (\ref{background}) from the perturbation variation. In Sec. \ref{sec: 3}, we write the second-order terms of the action with respect to the perturbations and derive the linearized field equations. In Sec. \ref{sec: 4}, we analyze the polarization modes of gravitational waves by using the linearized field equations. The second-order expansions of some important quantities on perturbations are sorted out in the Appendix.

\section{Background equations}
\label{sec: 2}
The Palatini formalism was first used by Einstein \cite{Einstein}, but it was named after Palatini \cite{Palatini1,Palatini2} for historical reasons. In this formalism, the connection is considered to be a variable independent of the metric. So we should take the variations of the gravitational action with respect to the metric and the connection independently. {The Riemann tensor $R^{\mu}_{\ \nu\rho\sigma}$ and Ricci tensor $R_{\mu\nu}$ in Palatini formalism are defined as}
\begin{eqnarray}
	{R^{\mu}_{\ \nu\rho\sigma}=\partial_{\rho}\Gamma^{\mu}_{\nu\sigma}-\partial_{\sigma}\Gamma^{\mu}_{\nu\rho}+\Gamma^{\mu}_{\lambda\rho}\Gamma^{\lambda}_{\nu\sigma}-\Gamma^{\mu}_{\lambda\sigma}\Gamma^{\lambda}_{\nu\rho},}
\\	
	\label{Ricci}
    R_{\mu\nu}=\partial_{\lambda}\Gamma^{\lambda}_{\mu\nu}-\partial_{\nu}\Gamma^{\lambda}_{\mu\lambda}+\Gamma^{\lambda}_{\sigma\lambda}\Gamma^{\sigma}_{\mu\nu}-\Gamma^{\lambda}_{\sigma\nu}\Gamma^{\sigma}_{\mu\lambda}.
\end{eqnarray}
For simplicity, we assume that the connection is nontorsion: $\Gamma^{\lambda}_{\mu\nu}=\Gamma^{\lambda}_{\nu\mu}$. In addition, it should be noted that the definition of $\tilde{\Box}\phi, \nabla^{\mu}\nabla^{\nu}\phi$ and $ \nabla^{\mu}\nabla_{\nu}\phi$ in action (\ref{action}) under Palatini formalism is not unique. This is because in Palatini formalism, the relationship $\nabla_{\lambda}g_{\mu\nu}=0$ is not generally satisfied \cite{Helpin}. In this paper, we define
\begin{eqnarray}	
\nonumber	\label{nabla define}
\tilde{\Box}\phi&=&g^{\mu\nu}\nabla_{\mu}\nabla_{\nu}\phi,
\\
\nabla^{\mu}\nabla^{\nu}\phi&=&g^{\mu\rho}\nabla_{\rho}\nabla^{\nu}\phi,
\\
\nabla^{\mu}\nabla_{\nu}\phi&=&g^{\mu\rho}\nabla_{\rho}\nabla_{\nu}\phi.\nonumber
\end{eqnarray}

If we use $\eta_{\mu\nu}$ and $\eta^{\mu\nu}$ to lower and raise the index of metric perturbation $h_{\mu\nu}$, and define $h=\eta^{\mu\nu}h_{\mu\nu}$, then using (\ref{perturbation})-(\ref{nabla define}) we can write out the first-order term of the perturbations in the action $S$:
\begin{eqnarray}
	\label{action 1 order}
	\nonumber
	\mathop{S}^{(1)} &=& \int d^4x
      \left\{  \frac{1}{2}h\left[\mathop{K}^{0}+\mathop{G}^{0}~\!\!^{}_{4}\eta^{\alpha\beta}\left(\mathop{\Gamma}^{0}~\!\!^{\lambda}_{\sigma\lambda}\mathop{\Gamma}^{0}~\!\!^{\sigma}_{\alpha\beta}-\mathop{\Gamma}^{0}~\!\!^{\lambda}_{\sigma\beta}\mathop{\Gamma}^{0}~\!\!^{\sigma}_{\alpha\lambda}\right)\right]
         \right. \\ \nonumber
         &+&\mathop{K}^{0}~\!\!^{}_{,\phi}\varphi-\mathop{G}^{0}~\!\!^{}_{3}\Box\varphi+\mathop{G}^{0}~\!\!^{}_{3}\eta^{\mu\nu}\mathop{\Gamma}^{0}~\!\!^{\lambda}_{\nu\mu}\partial_{\lambda}\varphi
         \\ \nonumber
         &+& \mathop{G}^{0}~\!\!^{}_{4}\eta^{\mu\nu}\left(\partial_{\lambda}\Sigma^{\lambda}_{\mu\nu}-\partial_{\nu}\Sigma^{\lambda}_{\mu\lambda}+\mathop{\Gamma}^{0}~\!\!^{\sigma}_{\mu\nu}\Sigma^{\lambda}_{\sigma\lambda}+\mathop{\Gamma}^{0}~\!\!^{\lambda}_{\sigma\lambda}\Sigma^{\sigma}_{\mu\nu}-\mathop{\Gamma}^{0}~\!\!^{\sigma}_{\mu\lambda}\Sigma^{\lambda}_{\sigma\nu}-\mathop{\Gamma}^{0}~\!\!^{\lambda}_{\sigma\nu}\Sigma^{\sigma}_{\mu\lambda}\right)
         \\ \nonumber
         &-&\mathop{G}^{0}~\!\!^{}_{4}h^{\mu\nu}\left(\mathop{\Gamma}^{0}~\!\!^{\lambda}_{\sigma\lambda}\mathop{\Gamma}^{0}~\!\!^{\sigma}_{\mu\nu}-\mathop{\Gamma}^{0}~\!\!^{\lambda}_{\sigma\nu}\mathop{\Gamma}^{0}~\!\!^{\sigma}_{\mu\lambda}\right)
         \\ \nonumber
         &+&\mathop{G}^{0}~\!\!^{}_{4,\phi}\varphi\eta^{\mu\nu}\left(\mathop{\Gamma}^{0}~\!\!^{\lambda}_{\sigma\lambda}\mathop{\Gamma}^{0}~\!\!^{\sigma}_{\mu\nu}-\mathop{\Gamma}^{0}~\!\!^{\lambda}_{\sigma\nu}\mathop{\Gamma}^{0}~\!\!^{\sigma}_{\mu\lambda}\right)
         \\ \nonumber
         &+&\mathop{G}^{0}~\!\!^{}_{5}\left[\mathop{\Gamma}^{0}~\!\!^{\lambda}_{\sigma\lambda}\mathop{\Gamma}^{0}~\!\!^{\sigma}_{\mu\nu}-\mathop{\Gamma}^{0}~\!\!^{\lambda}_{\sigma\nu}\mathop{\Gamma}^{0}~\!\!^{\sigma}_{\mu\lambda}-\frac{1}{2}\eta_{\mu\nu}\eta^{\alpha\beta}\left(\mathop{\Gamma}^{0}~\!\!^{\lambda}_{\sigma\lambda}\mathop{\Gamma}^{0}~\!\!^{\sigma}_{\alpha\beta}-\mathop{\Gamma}^{0}~\!\!^{\lambda}_{\sigma\beta}\mathop{\Gamma}^{0}~\!\!^{\sigma}_{\alpha\lambda}\right)\right]
         \\
         &\quad&\times
         \left. \left(\partial^{\mu}\partial^{\nu}\varphi
                      +\eta^{\mu\kappa}\mathop{\Gamma}^{0}~\!\!^{\nu}_{\delta\kappa}
                      \partial^{\delta}\varphi\right)
      \right\}.
\end{eqnarray}
Here, $\Box=\eta^{\mu\nu} \partial_{\mu}\partial_{\nu}$ and the {notation} ``0" above the letter means that the corresponding function takes the value of $(\phi = \phi_{0}, X = 0)$.

Varying the action (\ref{action 1 order}) with respect to $\varphi$ and $h^{\mu\nu}$, we obtain respectively
\begin{eqnarray}
\mathop{K}^{0}~\!\!^{}_{,\phi}+\mathop{G}^{0}~\!\!^{}_{4,\phi}\eta^{\mu\nu}
  \left(\mathop{\Gamma}^{0}~\!\!^{\lambda}_{\sigma\lambda}
        \mathop{\Gamma}^{0}~\!\!^{\sigma}_{\mu\nu}
       -\mathop{\Gamma}^{0}~\!\!^{\lambda}_{\sigma\nu}
        \mathop{\Gamma}^{0}~\!\!^{\sigma}_{\mu\lambda}\right)&=&0,
        \label{background equation phi} \\
   \frac{1}{2}\eta_{\mu\nu}\left[\mathop{K}^{0}
    +\mathop{G_{4}}^{0}\eta^{\alpha\beta}
    \left(\mathop{\Gamma}^{0}~\!\!^{\lambda}_{\sigma\lambda}
          \mathop{\Gamma}^{0}~\!\!^{\sigma}_{\alpha\beta}
         -\mathop{\Gamma}^{0}~\!\!^{\lambda}_{\sigma\beta}
          \mathop{\Gamma}^{0}~\!\!^{\sigma}_{\alpha\lambda}\right)\right] &&
    \nonumber \\
   -\mathop{G_{4}}^{0}\left(\mathop{\Gamma}^{0}~\!\!^{\lambda}_{\sigma\lambda}\mathop{\Gamma}^{0}~\!\!^{\sigma}_{\mu\nu}-\frac{1}{2}\mathop{\Gamma}^{0}~\!\!^{\lambda}_{\sigma\nu}\mathop{\Gamma}^{0}~\!\!^{\sigma}_{\mu\lambda}-\frac{1}{2}\mathop{\Gamma}^{0}~\!\!^{\lambda}_{\sigma\mu}
    \mathop{\Gamma}^{0}~\!\!^{\sigma}_{\nu\lambda}\right)&=&0.
   \label{background equation h}
\end{eqnarray}
By contracting Eq. (\ref{background equation h}) with $\eta^{\mu\nu}$, we have
\begin{eqnarray}
	\label{reduced background equation h} 2\mathop{K}^{0}+\mathop{G_{4}}^{0}\eta^{\alpha\beta}\left(\mathop{\Gamma}^{0}~\!\!^{\lambda}_{\sigma\lambda}\mathop{\Gamma}^{0}~\!\!^{\sigma}_{\alpha\beta}-\mathop{\Gamma}^{0}~\!\!^{\lambda}_{\sigma\beta}\mathop{\Gamma}^{0}~\!\!^{\sigma}_{\alpha\lambda}\right)=0.
\end{eqnarray}
Further, by substituting Eq. (\ref{reduced background equation h}) into Eq. (\ref{background equation h}), we can simplify Eq. (\ref{background equation h}) as
\begin{eqnarray} 	
	\label{simple background equation h}
     \frac{1}{2}\eta_{\mu\nu}\mathop{K}^{0}+\mathop{G_{4}}^{0}\left(\mathop{\Gamma}^{0}~\!\!^{\lambda}_{\sigma\lambda}\mathop{\Gamma}^{0}~\!\!^{\sigma}_{\mu\nu}-\frac{1}{2}\mathop{\Gamma}^{0}~\!\!^{\lambda}_{\sigma\nu}\mathop{\Gamma}^{0}~\!\!^{\sigma}_{\mu\lambda}-\frac{1}{2}\mathop{\Gamma}^{0}~\!\!^{\lambda}_{\sigma\mu}\mathop{\Gamma}^{0}~\!\!^{\sigma}_{\nu\lambda}\right)=0.
\end{eqnarray}

The field equation with respect to $\Sigma^{\lambda}_{\mu\nu}$ can also be obtained from the action (\ref{action 1 order}) as
\begin{eqnarray} 	
	\label{background equation sigma}
    \mathop{G_{4}}^{0}\left(\frac{1}{2}\eta^{\alpha\beta}\mathop{\Gamma}^{0}~\!\!^{\mu}_{\alpha\beta}\delta^{\nu}_{\lambda}+\frac{1}{2}\eta^{\alpha\beta}\mathop{\Gamma}^{0}~\!\!^{\nu}_{\alpha\beta}\delta^{\mu}_{\lambda}+\eta^{\mu\nu}\mathop{\Gamma}^{0}~\!\!^{\rho}_{\lambda\rho}-\eta^{\alpha\nu}\mathop{\Gamma}^{0}~\!\!^{\mu}_{\alpha\lambda}-\eta^{\alpha\mu}\mathop{\Gamma}^{0}~\!\!^{\nu}_{\alpha\lambda}\right)=0.
\end{eqnarray}
By contracting Eq. (\ref{background equation sigma}) with $\delta^{\lambda}_{\nu}$, we have
\begin{eqnarray} 	
	\label{reduced lambda nu background equation sigma} \frac{3}{2}\mathop{G_{4}}^{0}\eta^{\alpha\beta}\mathop{\Gamma}^{0}~\!\!^{\mu}_{\alpha\beta}=0,
\end{eqnarray}
with which and by contracting Eq. (\ref{background equation sigma}) with {$\eta_{\mu\nu}$}, we have
\begin{eqnarray} 	
	\label{reduced mu nu background equation sigma}
	2\mathop{G_{4}}^{0}\mathop{\Gamma}^{0}~\!\!^{\rho}_{\lambda\rho}=0.
\end{eqnarray}
So, Eq. (\ref{background equation sigma}) can be simplified to
\begin{eqnarray} 	
	\label{simple background equation sigma} \mathop{G_{4}}^{0}\left(\eta^{\alpha\nu}\mathop{\Gamma}^{0}~\!\!^{\mu}_{\alpha\lambda}+\eta^{\alpha\mu}\mathop{\Gamma}^{0}~\!\!^{\nu}_{\alpha\lambda}\right)=0.
\end{eqnarray}

Equations (\ref{background equation phi}), (\ref{simple background equation h}), and (\ref{simple background equation sigma}) are the required background field equations. We notice that when
\begin{eqnarray}
	\label{G4neq0}
	\mathop{G_{4}}^{0}\neq0,
\end{eqnarray}		
Eq. (\ref{simple background equation sigma}) implies
\begin{eqnarray} 	
	\label{almost Gamma=0}
	\eta^{\alpha\nu}\mathop{\Gamma}^{0}~\!\!^{\mu}_{\alpha\lambda}+\eta^{\alpha\mu}\mathop{\Gamma}^{0}~\!\!^{\nu}_{\alpha\lambda}=0.
\end{eqnarray}
After multiplying the above formula by $\eta_{\rho\mu}\eta_{\sigma\nu}$, we can rotate the {indices} $(\rho,\sigma,\lambda)$ to obtain three equations. When we subtract the third equation from the sum of any other two, we can get
\begin{eqnarray} 	
	\label{Gamma=0}	
	\mathop{\Gamma}^{0}~\!\!^{\lambda}_{\mu\nu}=0.
\end{eqnarray}
Further, by the use of Eqs. (\ref{background equation phi}) and (\ref{simple background equation h}), we have
\begin{eqnarray} 	
	\label{K,phi;K=0}
	\mathop{K}^{0}~\!\!^{}_{,\phi}=0,\quad\mathop{K}^{0}=0.
\end{eqnarray}
{It should be noted that Eq. (\ref*{K,phi;K=0}) only shows that $K(\phi,X)$ and its first-order partial derivative to $\phi$ are zero for the Minkowski background solution (\ref{background}). This does not imply that $K_{,X}, K_{,\phi\phi}$ and other derivatives of $K$ are also zero for this background solution.}
	
	%when taking the value of the background solution (\ref{background}), so there is no reason to think that they are also zero when taking other solutions. Furthermore, it is uncertain whether the other derivatives of $K(\phi,X)$ are zero at the background solution.}

\section{Linearized field equations}
\label{sec: 3}
In this section, we will derive the linearized field equations of the perturbations.

When $G_{4}(\phi_{0},0)\neq0$, using (\ref{Gamma=0}) and (\ref{K,phi;K=0}), we can expand the second-order terms of the perturbations in the action (\ref{action}) as
\begin{eqnarray} 	
	\mathop{S}^{(2)} &=& \int d^4x
\left\{
   -\frac{1}{2}\mathop{K}^{0}~\!\!^{}_{,X}\partial^{\mu}\varphi\partial_{\mu}\varphi+\frac{1}{2}\mathop{K}^{0}~\!\!^{}_{\phi\phi}\varphi^{2}
	\right. \nonumber \\ \nonumber	&+&\mathop{G_{3}}^{0}\left(\eta^{\mu\nu}\Sigma^{\lambda}_{\mu\nu}\partial_{\lambda}\varphi+h^{\mu\nu}\partial_{\mu}\partial_{\nu}\varphi\right)-\mathop{G}^{0}~\!\!^{}_{3,\phi}\varphi\Box\varphi-\frac{1}{2}\mathop{G_{3}}^{0}h\Box\varphi
	\\ \nonumber	&+&\mathop{G_{4}}^{0}\eta^{\mu\nu}\left(\Sigma^{\lambda}_{\sigma\lambda}\Sigma^{\sigma}_{\mu\nu}-\Sigma^{\lambda}_{\sigma\nu}\Sigma^{\sigma}_{\mu\lambda}\right)
    -\mathop{G_{4}}^{0}h^{\mu\nu}\left(\partial_{\lambda}\Sigma^{\lambda}_{\mu\nu}-\partial_{\nu}\Sigma^{\lambda}_{\mu\lambda}\right)
	\\ \nonumber	&+&\mathop{G}^{0}~\!\!^{}_{4,\phi}\varphi\eta^{\mu\nu}\left(\partial_{\lambda}\Sigma^{\lambda}_{\mu\nu}-\partial_{\nu}\Sigma^{\lambda}_{\mu\lambda}\right)
   +\frac{1}{2}\mathop{G_{4}}^{0}h\eta^{\mu\nu}\left(\partial_{\lambda}\Sigma^{\lambda}_{\mu\nu}-\partial_{\nu}\Sigma^{\lambda}_{\mu\lambda}\right)
	\\ \nonumber	&+&\mathop{G}^{0}~\!\!^{}_{4,X}\left[(\Box\varphi)^{2}-\partial_{\mu}\partial_{\nu}\varphi\partial^{\mu}\partial^{\nu}\varphi\right]
	\\	&+&
   \left. \mathop{G_{5}}^{0}
       \left[\partial_{\lambda}\Sigma^{\lambda}_{\mu\nu}
             -\partial_{\nu}\Sigma^{\lambda}_{\mu\lambda}
             -\frac{1}{2}\eta_{\mu\nu}\eta^{\alpha\beta}
               \left(\partial_{\lambda}\Sigma^{\lambda}_{\alpha\beta}
                     -\partial_{\beta}\Sigma^{\lambda}_{\alpha\lambda}
               \right)
       \right]
    \partial^{\mu}\partial^{\nu}\varphi
    \right\}.
\label{action 2 order}	
\end{eqnarray}

Varying the action (\ref{action 2 order}) with respect to $h^{\mu\nu}$, we obtain
\begin{eqnarray} 	
	\label{linearized equcation h}
	\nonumber
		&&\mathop{G_{3}}^{0}\left(\partial_{\mu}\partial_{\nu}\varphi-\frac{1}{2}\eta_{\mu\nu}\Box\varphi\right)-\mathop{G_{4}}^{0}\left(\partial_{\lambda}\Sigma^{\lambda}_{\mu\nu}-\frac{1}{2}\partial_{\nu}\Sigma^{\lambda}_{\mu\lambda}-\frac{1}{2}\partial_{\mu}\Sigma^{\lambda}_{\nu\lambda}\right)
		\\
		&\quad&\quad\qquad\qquad+\frac{1}{2}\mathop{G_{4}}^{0}\eta_{\mu\nu}\eta^{\alpha\beta}\left(\partial_{\lambda}\Sigma^{\lambda}_{\alpha\beta}-\partial_{\beta}\Sigma^{\lambda}_{\alpha\lambda}\right)=0.
\end{eqnarray}
By contracting Eq. (\ref{linearized equcation h}) with $\eta^{\mu\nu}$, we have
\begin{eqnarray} 	
	\label{linearized equcation h reduce munu}
	-\mathop{G_{3}}^{0}\Box\varphi+\mathop{G_{4}}^{0}\eta^{\alpha\beta}\left(\partial_{\lambda}\Sigma^{\lambda}_{\alpha\beta}-\partial_{\beta}\Sigma^{\lambda}_{\alpha\lambda}\right)=0.
\end{eqnarray}
Then, substituting Eq. (\ref*{linearized equcation h reduce munu}) into Eq. (\ref{linearized equcation h}) yields
\begin{eqnarray} 	
	\label{simple linearized equcation h}
	\mathop{G_{3}}^{0}\partial_{\mu}\partial_{\nu}\varphi-\mathop{G_{4}}^{0}\left(\partial_{\lambda}\Sigma^{\lambda}_{\mu\nu}-\frac{1}{2}\partial_{\nu}\Sigma^{\lambda}_{\mu\lambda}-\frac{1}{2}\partial_{\mu}\Sigma^{\lambda}_{\nu\lambda}\right)=0.
\end{eqnarray}

The linear field equation for $\varphi$ derived from (\ref{action 2 order}) is
\begin{eqnarray}
	\label{linearized equcation phi}
	\nonumber
	&&\qquad\left(\mathop{K}^{0}~\!\!^{}_{,X}-2	\mathop{G}^{0}~\!\!^{}_{3,\phi}\right)\Box\varphi+\mathop{K}^{0}~\!\!^{}_{,\phi\phi}\varphi+	\mathop{G_{3}}^{0}\left(\partial_{\mu}\partial_{\nu}h^{\mu\nu}-\frac{1}{2}\Box h\right)
	\\ \nonumber
	&&\quad\qquad-\mathop{G_{3}}^{0}\eta^{\mu\nu}\partial_{\lambda}\Sigma^{\lambda}_{\mu\nu}	+\mathop{G}^{0}~\!\!^{}_{4,\phi}\eta^{\mu\nu}\left(\partial_{\lambda}\Sigma^{\lambda}_{\mu\nu}-\partial_{\nu}\Sigma^{\lambda}_{\mu\lambda}\right)
	\\
	&+&\mathop{G_{5}}^{0}\partial^{\mu}\partial^{\nu}\left[\partial_{\lambda}\Sigma^{\lambda}_{\mu\nu}-\partial_{\nu}\Sigma^{\lambda}_{\mu\lambda}-\frac{1}{2}\eta_{\mu\nu}\eta^{\alpha\beta}\left(\partial_{\lambda}\Sigma^{\lambda}_{\alpha\beta}-\partial_{\beta}\Sigma^{\lambda}_{\alpha\lambda}\right)\right]=0.
\end{eqnarray}
Substituting Eq. (\ref{simple linearized equcation h}) into Eq. (\ref{linearized equcation phi}), we get
\begin{eqnarray}
	\label{simple linearized equcation phi}
	\nonumber
    \left(\mathop{K}^{0}~\!\!^{}_{,X}-2\mathop{G}^{0}~\!\!^{}_{3,\phi}\right)\Box\varphi+\mathop{K}^{0}~\!\!^{}_{,\phi\phi}\varphi+\mathop{G_{3}}^{0}\left(\partial_{\mu}\partial_{\nu}h^{\mu\nu}-\frac{1}{2}\Box h\right)
    &-&\mathop{G_{3}}^{0}\eta^{\mu\nu}\partial_{\lambda}\Sigma^{\lambda}_{\mu\nu}
	\\
    +\left(\mathop{G}^{0}~\!\!^{}_{4,\phi}\mathop{G_{3}}^{0}/\mathop{G_{4}}^{0}\right)
    \Box\varphi+\frac{1}{2}
    \left(\mathop{G_{5}}^{0}\mathop{G_{3}}^{0}/\mathop{G_{4}}^{0}\right)
    \Box\Box\varphi&=&0.
\end{eqnarray}
Note that the above field equation for the perturbations is {fourth-order}, which is very different from the metric Horndeski theory.

At last, varying the action (\ref{action 2 order}) with respect to $\Sigma^{\lambda}_{\mu\nu}$, we get
\begin{eqnarray}
	\label{linearized equcation Sigma}
	\nonumber
	&\quad&\mathop{G_{3}}^{0}\eta^{\mu\nu}\partial_{\lambda}\varphi-\mathop{G}^{0}~\!\!^{}_{4,\phi}\left(\eta^{\mu\nu}\partial_{\lambda}\varphi-\frac{1}{2}\delta^{\nu}_{\lambda}\partial^{\mu}\varphi-\frac{1}{2}\delta^{\mu}_{\lambda}\partial^{\nu}\varphi\right)
	\\ \nonumber
    &-&\mathop{G_{5}}^{0}\left(\partial_{\lambda}\partial^{\mu}\partial^{\nu}\varphi-\frac{1}{2}\delta^{\nu}_{\lambda}\partial^{\mu}\Box\varphi-\frac{1}{2}\delta^{\mu}_{\lambda}\partial^{\nu}\Box\varphi\right)
    \\ \nonumber
    &+&\frac{1}{2}\mathop{G_{5}}^{0}\left(\eta^{\mu\nu}\partial_{\lambda}\Box\varphi-\frac{1}{2}\delta^{\nu}_{\lambda}\partial^{\mu}\Box\varphi-\frac{1}{2}\delta^{\mu}_{\lambda}\partial^{\nu}\Box\varphi\right)
    \\ \nonumber
    &+&\mathop{G_{4}}^{0}\left[\partial_{\lambda}h^{\mu\nu}-\frac{1}{2}\delta^{\nu}_{\lambda}\partial_{\rho}h^{\mu\rho}-\frac{1}{2}\delta^{\mu}_{\lambda}\partial_{\rho}h^{\nu\rho}-\frac{1}{2}\left(\eta^{\mu\nu}\partial_{\lambda}h-\frac{1}{2}\delta^{\nu}_{\lambda}\partial^{\mu}h-\frac{1}{2}\delta^{\mu}_{\lambda}\partial^{\nu}h\right)\right]
    \\
    &+&\mathop{G_{4}}^{0}\left(\eta^{\mu\nu}\Sigma^{\rho}_{\lambda\rho}+\frac{1}{2}\delta^{\nu}_{\lambda}\eta^{\alpha\beta}\Sigma^{\mu}_{\alpha\beta}+\frac{1}{2}\delta^{\mu}_{\lambda}\eta^{\alpha\beta}\Sigma^{\nu}_{\alpha\beta}-\eta^{\nu\rho}\Sigma^{\mu}_{\rho\lambda}-\eta^{\mu\rho}\Sigma^{\nu}_{\rho\lambda}\right)=0.
\end{eqnarray}
By contracting Eq. (\ref{linearized equcation Sigma}) with $\delta^{\lambda}_{\mu}$, we have
\begin{eqnarray} 	
	\label{linearized equcation Sigma reduce mulambda}
	\frac{2}{3}\mathop{G_{3}}^{0}\partial^{\nu}\varphi+\mathop{G}^{0}~\!\!^{}_{4,\phi}\partial^{\nu}\varphi+\frac{1}{2}\mathop{G_{5}}^{0}\partial^{\nu}\Box\varphi-\mathop{G_{4}}^{0}\left(\partial_{\lambda}h^{\lambda\nu}-\frac{1}{2}\partial^{\nu}h\right)+\mathop{G_{4}}^{0}\eta^{\lambda\rho}\Sigma^{\nu}_{\rho\lambda}=0.
\end{eqnarray}
Substituting Eq. (\ref{linearized equcation Sigma reduce mulambda}) into Eq. (\ref{linearized equcation Sigma}), and contracting Eq. (\ref{linearized equcation Sigma}) with $\eta^{\mu\nu}$, we get the following equation
\begin{eqnarray} 	
	\label{linearized equcation Sigma reduce munu}
	\frac{5}{3}\mathop{G_{3}}^{0}\partial_{\lambda}\varphi-2\mathop{G}^{0}~\!\!^{}_{4,\phi}\partial_{\lambda}\varphi+\frac{1}{2}\mathop{G_{5}}^{0}\Box\partial_{\lambda}\varphi-\frac{1}{2}\mathop{G_{4}}^{0}\partial_{\lambda}h+\mathop{G_{4}}^{0}\Sigma^{\rho}_{\lambda\rho}=0.
\end{eqnarray}
Using Eqs. (\ref{linearized equcation Sigma reduce mulambda}) and  (\ref{linearized equcation Sigma reduce munu}), we can simplify Eq. (\ref{linearized equcation Sigma}) as
\begin{eqnarray}
	\label{simple linearized equcation Sigma}
	\nonumber	\mathop{G_{4}}^{0}\left(\eta^{\nu\rho}\Sigma^{\mu}_{\rho\lambda}+\eta^{\mu\rho}\Sigma^{\nu}_{\rho\lambda}\right)&=&-\frac{2}{3}\mathop{G_{3}}^{0}\eta^{\mu\nu}\partial_{\lambda}\varphi
    -\frac{1}{3}\mathop{G_{3}}^{0}\delta^{\mu}_{\lambda}\partial^{\nu}\varphi-\frac{1}{3}\mathop{G_{3}}^{0}\delta^{\nu}_{\lambda}\partial^{\mu}\varphi
    \\
    &+&\mathop{G}^{0}~\!\!^{}_{4,\phi}\eta^{\mu\nu}\partial_{\lambda}\varphi-\mathop{G_{5}}^{0}\partial_{\lambda}\partial^{\mu}\partial^{\nu}\varphi+\mathop{G_{4}}^{0}\partial_{\lambda}h^{\mu\nu}.
\end{eqnarray}
Similar to the process of getting Eq. (\ref{Gamma=0}) from Eq. (\ref{almost Gamma=0}), we can use Eq. (\ref*{simple linearized equcation Sigma}) to get
\begin{eqnarray}
	\label{Sigma=}
	\nonumber
	\Sigma^{\lambda}_{\mu\nu}&=&\frac{1}{2}\left(\mathop{G}^{0}~\!\!^{}_{4,\phi}/\mathop{G_{4}}^{0}\right)\left(\delta^{\lambda}_{\mu}\partial_{\nu}\varphi+\delta^{\lambda}_{\nu}\partial_{\mu}\varphi-\eta_{\mu\nu}\partial^{\lambda}\varphi\right)
	\\ \nonumber
	&-&\frac{1}{3}\left(\mathop{G_{3}}^{0}/\mathop{G_{4}}^{0}\right)\left(\delta^{\lambda}_{\mu}\partial_{\nu}\varphi+\delta^{\lambda}_{\nu}\partial_{\mu}\varphi\right)
	\\ \nonumber
	&-& \frac{1}{2}\left(\mathop{G_{5}}^{0}/\mathop{G_{4}}^{0}\right)\partial^{\lambda}\partial_{\mu}\partial_{\nu}\varphi
	\\
	&+&\left[\frac{1}{2}\eta^{\lambda\sigma}\left(\partial_{\nu}h_{\sigma\mu}+\partial_{\mu}h_{\nu\sigma}-\partial_{\sigma}h_{\mu\nu}\right)\right],
\end{eqnarray}
where the quantity in square bracket is the Levi-Civita connection. Substituting (\ref{Sigma=}) into Eqs. (\ref{simple linearized equcation h}) and (\ref{simple linearized equcation phi}), we get the final equations
\begin{eqnarray}
	\label{R_munu=}
    \mathop{R_{\mu\nu}}^{{(LC)}~~}=\left(\mathop{G}^{0}~\!\!^{}_{4,\phi}/\mathop{G_{4}}^{0}\right)\left(\partial_{\mu}\partial_{\nu}\varphi+\frac{1}{2}\eta_{\mu\nu}\Box\varphi\right),
\\
 \label{varphi=}
	a\Box\Box\varphi+b\Box\varphi+c\varphi=0.\qquad\qquad
\end{eqnarray}
Here, the {notation $(LC)$} above $R_{\mu\nu}$ indicates that this is the Ricci tensor defined by the Levi-Civita connection. {So it is the same as the Ricci tensor in the metric formalism and is different from Eq. (\ref{Ricci}). In addition,}
\begin{eqnarray}
	\label{a,b,c}
	\nonumber
a&=&\mathop{G_{5}}^{0}\mathop{G_{3}}^{0}/\mathop{G_{4}}^{0},
		\\ \nonumber		b&=&\mathop{K}^{0}~\!\!^{}_{,X}-2\mathop{G}^{0}~\!\!^{}_{3,\phi}
  +2\mathop{G}^{0}~\!\!^{}_{4,\phi}\mathop{G_{3}}^{0}/\mathop{G_{4}}^{0}
  +\frac{2}{3}\left(\mathop{G_{3}}^{0}\right)^{2}/\mathop{G_{4}}^{0},
		\\
		c&=&\mathop{K}^{0}~\!\!^{}_{,\phi\phi}.
\end{eqnarray}

Now we consider the case of $G_{4}(\phi_{0},0)=0$. When we expand the second-order terms of the perturbations in the action ($\ref{action}$), and vary the action with respect to $h^{\mu\nu}$ and $\Sigma^{\lambda}_{\mu\nu}$, we find that all the equations are these satisfied by $\varphi$ in this case. The field equations of $h_{\mu\nu}$ and $\Sigma^{\lambda}_{\mu\nu}$ can only be obtained by varying the action ($\ref{action}$) with respect to $\varphi$. However, this means that the independent components contained in $h_{\mu\nu}$ and $\Sigma^{\lambda}_{\mu\nu}$ are limited by only one equation. So we exclude the case of $G_{4}(\phi_{0},0)=0$ in the following discussions.

\section{Polarization modes of gravitational waves}
\label{sec: 4}
A framework for analyzing the polarization modes of gravitational waves propagating at the speed of light in the Newman-Penrose formalism was given in Ref. \cite{Eardley}. And in Ref.  \cite{TH. Hyun}, this framework was extended to the case of gravitational waves propagating at arbitrary speed. In this section, we first summarize the Newman-Penrose formalism of gravitational waves propagating at arbitrary speed, and then analyze the polarization modes of gravitational waves in Palatini-Horndeski gravity {with this method. It should be pointed out that in addition to the Newman-Penrose formalism, a gauge invariant method can also be used to analyze the propagation degrees of freedom and polarization modes of gravitational waves \cite{TeVeS,Fabio Moretti 3}.} All Riemannian tensor $R_{\mu\nu\lambda\rho}$, Ricci tensor $R_{\mu\nu}$ and Ricci scalar $R$ that appear in this section are defined by the Levi-Civita connection, so we omit the {notation $(LC)$} on these quantities.

Because the action describing the motion of a free particle in Palatini-Horndeski theory is still
\begin{eqnarray}
	\label{free particle action}
    S=\int ds,
\end{eqnarray}
the relative motion between two adjacent test particles used to detect the polarization modes of gravitational waves still satisfies the equation of geodesic deviation \cite{MTW}:
\begin{eqnarray}
	\label{equation of geodesic deviation}
	\frac{d^{2}\eta_{i}}{dt^{2}}=-R_{i0j0}\eta^{j}.
\end{eqnarray}
Here, $\eta_{i}$ is the relative displacement of the two test particles. The Latin alphabet indices $(i,j,k)$ range over space indices ($1,2,3$) which point to ($+x,+y,+z$) directions respectively. When the superscript and subscript are the same Latin alphabet, sum the space index. It can be seen from Eq. (\ref*{equation of geodesic deviation}) that $R_{i0j0}$ will completely determine the motion behavior of the test particles under the gravitational waves, and the polarization modes of gravitational waves are defined by different values of $R_{i0j0}$ \cite{MTW}.

Now take the propagation direction of gravitational waves as $+z$ direction. Mark the components of $R_{i0j0}$ as ($i$ is row index and $j$ column one)
\begin{eqnarray}
	\label{P1-P6}
	R_{i0j0}=\begin{pmatrix}
		P_{4}+P_{6} & P_{5} & P_{2}\\
		P_{5}       & -P_{4}+P_{6}  & P_{3}\\
		P_{2}       &  P_{3}   &   P_{1}
	\end{pmatrix}.
\end{eqnarray}
The six independent variables, $P_{1},...,P_{6}$ can define six basic polarization modes according to the geodesic deviation equation (\ref{equation of geodesic deviation}). Taking $P_{1},...,P_{6}$ as the monochromatic plane wave solutions and bring (\ref*{P1-P6}) into Eq. (\ref*{equation of geodesic deviation}), one can show the relative motions of the test particles under these six polarization modes in Fig. 1.

\begin{figure*}[htbp]
	\makebox[\textwidth][c]{\includegraphics[width=1.2\textwidth]{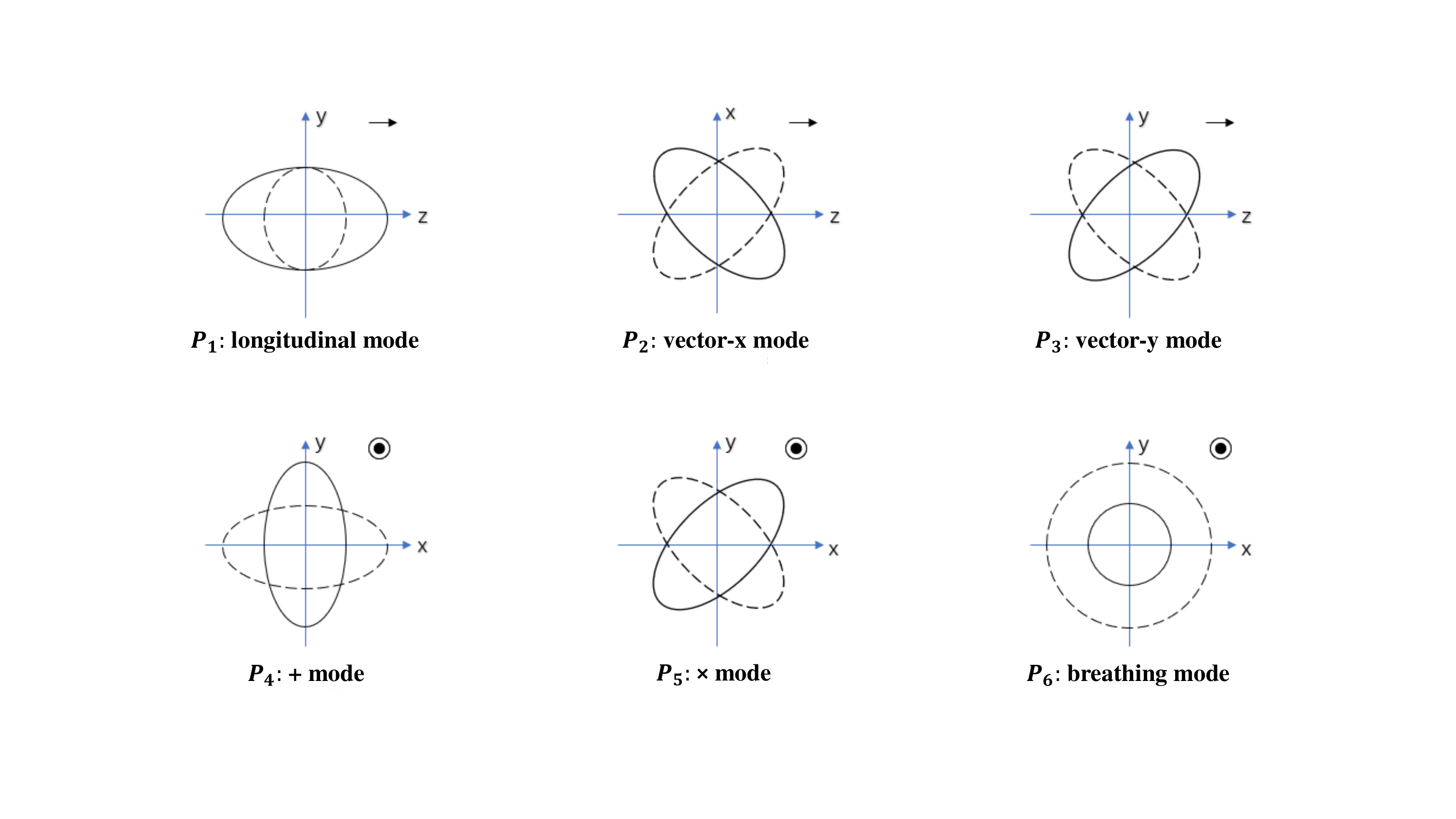}}
	\caption{Six polarization modes of gravitational waves \cite{Eardley}. The gravitational waves in the figure propagate in the $+z$ direction. The solid line is the case of a circle of test particles when the phase of waves is $0$, and the dotted line is the case of a circle of test particles when the phase is $\pi$. There is no relative motion between test particles in the direction of the third axis that is not drawn.}
	\label{fig: 1}
\end{figure*}

Now consider that the Riemannian tensor only depends on the variable $u=t-z/v$ ($v$ is the wave speed). When taking the null tetrad basis representation of Newman-Penrose formalism \cite{N-P}:
\begin{eqnarray}
	\label{N-P base k}
	k^{\mu}&=&\frac{1}{\sqrt{2}}\left(1,0,0,1\right),\\
	\label{N-P base l}
	l^{\mu}&=&\frac{1}{\sqrt{2}}\left(1,0,0,-1\right),\\
	\label{N-P base m}
	m^{\mu}&=&\frac{1}{\sqrt{2}}\left(0,1,i,0\right),\\
	\label{N-P base m bar}
	\bar{m}^{\mu}&=&\frac{1}{\sqrt{2}}\left(0,1,-i,0\right).
\end{eqnarray}
We can find $\nabla_{\mu} u$ will be a linear combination of $k_{\mu}$ and $l_{\mu}$. So the Riemannian tensor that depends only on the variable $u$ satisfies
\begin{eqnarray}
	\label{Rabcd,p=0}
	R_{abcd,p}=0.
\end{eqnarray}
Here the indices $(a,b,c,d)$ range over $(k,l,m,\bar{m})$, and the indices $(p,q,r)$ range over $(m,\bar{m})$. Further use the Bianchi identity, we get
\begin{eqnarray}
	\label{Rabpq,c=0}
	R_{ab[pq,c]}=R_{abpq,c}=0.
\end{eqnarray}
So the nonvanishing components of the Riemannian tensor (except a constant independent of fluctuations) include
\begin{eqnarray}
	\label{non-zero component of Riemannian tensor}
	\nonumber
	R_{klkl},\quad R_{klml},\quad R_{kl\bar{m}l},\quad R_{klmk},\quad R_{kl\bar{m}k},\quad R_{mlml},\quad R_{ml\bar{m}l},\quad R_{mlmk},
	\\
	R_{ml\bar{m}k},\quad R_{\bar{m}l\bar{m}l},\quad R_{\bar{m}l\bar{m}k},\quad R_{\bar{m}lmk},\quad R_{mkmk},\quad R_{mk\bar{m}k},\quad R_{\bar{m}k\bar{m}k}.
\end{eqnarray}
Writing the components of $R_{i0j0}$ under the base (\ref{N-P base k} - \ref{N-P base m bar}), we can get
\begin{eqnarray}
	\label{P1-P6=}
	\nonumber
	P_{1}&=&R_{klkl},
	\\ \nonumber
	P_{2}&=&\frac{1}{2}\left(R_{mlkl}+R_{\bar{m}lkl}\right)-\frac{1}{2}\left(R_{mklk}+R_{\bar{m}klk}\right),
	\\ \nonumber
	P_{3}&=&-\frac{i}{2}\left(R_{mlkl}-R_{\bar{m}lkl}\right)+\frac{i}{2}\left(R_{mklk}-R_{\bar{m}klk}\right),
	\\ \nonumber
	P_{4}&=&\frac{1}{4}\left(R_{mkmk}+R_{\bar{m}k\bar{m}k}\right)+\frac{1}{4}\left(R_{mlml}+R_{\bar{m}l\bar{m}l}\right)+\frac{1}{2}\left(R_{mkml}+R_{\bar{m}k\bar{m}l}\right),
	\\ \nonumber
	P_{5}&=&-\frac{i}{4}\left(R_{mkmk}-R_{\bar{m}k\bar{m}k}\right)-\frac{i}{4}\left(R_{mlml}-R_{\bar{m}l\bar{m}l}\right)-\frac{i}{2}\left(R_{mkml}-R_{\bar{m}k\bar{m}l}\right),
	\\
	P_{6}&=&\frac{1}{2}\left(R_{mk\bar{m}l}+R_{\bar{m}kml}\right)+\frac{1}{2}R_{mk\bar{m}k}+\frac{1}{2}R_{ml\bar{m}l}.
\end{eqnarray}
Finally, the components of the Ricci tensor will become
\begin{eqnarray}
	\label{Ricci in N-P}
	\nonumber
	R_{ll}=2R_{ml\bar{m}l},\quad R_{lk}=R_{lmk\bar{m}}+R_{l\bar{m}km}+R_{lklk},\quad R_{kk}=2R_{kmk\bar{m}},\quad
	\\
	R_{lm}=R_{lklm},\quad R_{km}=R_{klkm},\quad R_{mm}=-2R_{mlmk},\quad R_{m\bar{m}}=-2R_{ml\bar{m}k}.
\end{eqnarray}
There are some components here that are not written. Some of them are complex conjugates of the others in (\ref{Ricci in N-P}). These components can be obtained by changing the indices $m$ and $\bar{m}$ into $\bar{m}$ and $m$, respectively. Others are zero (except a constant independent of fluctuations).

For example, for the vacuum field equation in general relativity $R_{\mu\nu}=0$, by using (\ref*{P1-P6=}) and (\ref*{Ricci in N-P}), we can find that only $P_{4}$ and $P_{5}$ cannot be zero. Therefore, general relativity only allows $+$ and $\times$ modes, which is consistent with the known result \cite{MTW}.

Now come back to analyze the polarization modes of gravitational waves in Palatini-Horndeski gravity. According to the investigation in the previous section, only the case of $G_{4}(\phi_{0},0)\neq0$ needs to be considered.

If $G_{4,\phi}(\phi_{0},0)=0$, Eq. (\ref*{R_munu=}) will become
\begin{eqnarray} 	
	\label{GR Equcation}
	{R_{\mu\nu}=\frac{1}{2}\left(\partial_{\sigma}\partial_{\nu}h^{\sigma}_{\mu}+\partial_{\sigma}\partial_{\mu}h^{\sigma}_{\nu}-\partial_{\mu}\partial_{\nu}h-\Box h_{\mu\nu}\right)=0.}
\end{eqnarray}
{After defining $\bar{h}_{\mu\nu}=h_{\mu\nu}-\frac{1}{2}\eta_{\mu\nu}h$ and taking Lorentz gauge $\partial_{\mu}\bar{h}^{\mu\nu}=0$, Eq. (\ref{GR Equcation}) becomes $\Box\bar{h}_{\mu\nu}=0$. In this case, the tensor perturbation and scalar perturbation are decoupled. This is the same as the case in general relativity. The solutions of $h_{\mu\nu}$ are plane waves, which only allow $+$ and $\times$ modes propagating at the speed of light. These wave solutions do not diverge with the time $t$, so they are linearly stable.}

% $R_{\mu\nu}=\textcolor{blue}{\frac{1}{2}(\partial_{\sigma}\partial_{\nu}h^{\sigma}_{\mu}+\partial_{\sigma}\partial_{\mu}h^{\sigma}_{\nu}-\partial_{\mu}\partial_{\nu}h-\Box h_{\mu\nu})=0$. In this case, only $+$ and $\times$ modes propagating at the speed of light are allowed. This is the same as in general relativity.}

So we just need to consider the case of $G_{4,\phi}(\phi_{0},0)\neq0$. Before doing that, it is convenient to consider the case that $\varphi$ has the following solution
\begin{eqnarray} 	
	\label{varphi=e^ikx}
	\varphi=\varphi_{0} e^{ikx},\qquad k^{2}=-m^2.
\end{eqnarray}
Here $\varphi_{0}$ {is the amplitude of the scalar field and $m$ is effective mass, both of which are constants.} The space component of $k^{\mu}$ points in the $+z$ direction. Then Eq. (\ref*{R_munu=}) becomes
\begin{eqnarray} 	
	\label{R_munu=e^ikx}
	\nonumber
	R_{\mu\nu}
	    &=&{\frac{1}{2}\left(\partial_{\sigma}\partial_{\nu}h^{\sigma}_{\mu}+\partial_{\sigma}\partial_{\mu}h^{\sigma}_{\nu}-\partial_{\mu}\partial_{\nu}h-\Box h_{\mu\nu}\right)}
	    \\
		&=&-\left(\varphi_{0} \mathop{G}^{0}~\!\!^{}_{4,\phi}/\mathop{G_{4}}^{0}\right)\left(k_{\mu}k_{\nu}e^{ikx}-\frac{1}{2}\eta_{\mu\nu}m^{2}e^{ikx}\right).
\end{eqnarray}
{Since the nonzero components of the wave vector $k_{\mu}$ are $k_0$ and $k_3$, we have $k_m=k_{\bar{m}}=0$ under the Newman-Penrose formalism. This is because the bases ($m, \bar{m}$) are in the $x-y$ plane. In addition, the other two bases $(k, l)$ of the Newman-Penrose formalism are} {null vectors} {in the $t-z$ plane. Under the Newman-Penrose formalism, 1)  for a lightlike wave vector $k_{\mu}$, one has $k_k=0$ and $k_l \neq 0$. 2) for a nonlightlike wave vector $k_{\mu}$, one has $k_k\neq0$ and $k_l \neq 0$. Using these results, we can get the nonvanishing components of the Ricci tensor under the base (\ref{N-P base k} - \ref{N-P base m bar}): }%So the components of the Ricci tensor under the base (\ref{N-P base k} - \ref{N-P base m bar}) become
\begin{eqnarray}
	 	\label{Ricci = e^ikx}
	\nonumber
	R_{ll}&=&-\left(\varphi_{0}
	 \mathop{G}^{0}~\!\!^{}_{4,\phi}/\mathop{G_{4}}^{0}\right)k_{l}k_{l}e^{ikx},
	\\ \nonumber
	 R_{lk}&=&-\left(\varphi_{0} \mathop{G}^{0}~\!\!^{}_{4,\phi}/\mathop{G_{4}}^{0}\right)\left(k_{l}k_{k}e^{ikx}+\frac{1}{2}m^{2}e^{ikx}\right),
	 \\ \nonumber
	 R_{kk}&=&-\left(\varphi_{0} \mathop{G}^{0}~\!\!^{}_{4,\phi}/\mathop{G_{4}}^{0}\right)k_{k}k_{k}e^{ikx},
	 \\ \nonumber
	 R_{m\bar{m}}&=&{\frac{1}{2}\left(\varphi_{0}\mathop{G}^{0}~\!\!^{}_{4,\phi}/\mathop{G_{4}}^{0}\right) m^{2}e^{ikx},}\\
	R_{lm}&=&{0,\quad R_{km}=0,\quad R_{mm}=0.}
\end{eqnarray}
Comparing Eqs. (\ref{Ricci = e^ikx}) and (\ref{Ricci in N-P}), and using Eq. (\ref{P1-P6=}), we can show that $P_{2}$ and $P_{3}$ must be zero {and $P_{4}$ and $P_{5}$ cannot be zero. This means that there must be no vector polarization modes and it allows $+$ and $\times$ modes to propagate at the speed of light in Palatini-Horndeski theory.} In addition, 1) when $m^{2}$ is zero, besides $P_{4}$ and $P_{5}$, only $P_{6}$ cannot be zero. {At this point, it also allows} a breathing mode propagating at the speed of light, corresponding to a massless scalar degree of freedom. And 2) when $m^{2}$ is not zero, besides $P_{4}$ and $P_{5}$, $P_{1}$ and $P_{6}$ are all not zero. For this case, we {also have a mixture mode} of the breathing and longitudinal modes propagating at the speed less than the speed of light. The mixture mode corresponds to a massive scalar degree of freedom.

{Further, the general solution of Eq. (\ref{R_munu=e^ikx}) has the following form}
\begin{eqnarray}
	\label{solution Eq. (52)}
	{h_{\mu\nu}=A_{\mu\nu}e^{ikx}+f_{\mu\nu}(x^{\mu})},
\end{eqnarray}
{where}
\begin{eqnarray}
	\label{A=}
	\nonumber
	{A_{11}=A_{22}=-\varphi_{0} \mathop{G}^{0}~\!\!^{}_{4,\phi}/\mathop{G_{4}}^{0}, A_{33}=-\frac{m^2}{{k_{0}}^{2}}\varphi_{0} \mathop{G}^{0}~\!\!^{}_{4,\phi}/\mathop{G_{4}}^{0},}
	\\
  {A_{00}=A_{01}=A_{02}=A_{03}=A_{12}=A_{13}=A_{23}=0,}
\end{eqnarray}
{and $f_{\mu\nu}(x^{\mu})$ is the general solution of Eq. (\ref{GR Equcation}). We already know that $f_{\mu\nu}(x^{\mu})$ is linearly stable. It corresponds to tensor mode propagating at the speed of light. For $A_{\mu\nu}e^{ikx}$, it corresponds to a scalar mode with mass $m$. When $\varphi$ is a plane wave solution, $A_{\mu\nu}e^{ikx}$ is also a plane wave solution, so it is linearly stable.}

Next, we analyze the polarization modes of gravitational waves in the case of $G_{4,\phi}(\phi_{0},0)\neq0$. In the following discussion, we let the propagation direction of the waves be the $+z$ direction and let $k_{0}\textless0$.

We first consider Eq. (\ref{varphi=}) satisfied by $\varphi$ with $a=0$:

\textbf{\emph{Case 1.1}} $a=b=c=0$. There is no equation to constrain the scalar field $\varphi$ { under linear approximation. The equation of the scalar field perturbation $\varphi$ is at least quadratic, so we do not consider this case.	}
%, which is not in line with the physical reality, so it is excluded.

\textbf{\emph{Case 1.2}} $a=b=0, c\neq0$. Equation (\ref{varphi=}) tells us $\varphi=0$, and then Eq. (\ref{R_munu=}) become $R_{\mu\nu}=0$. {This solution is linearly stable.} In this case, only $+$ and $\times$ modes propagating at the speed of light are allowed.

\textbf{\emph{Case 1.3}} $a=c=0, b\neq0$. Equation (\ref{varphi=}) becomes $\Box\varphi=0$, then $\varphi
$ has a solution
\begin{eqnarray} 	
	\varphi=\varphi_{0} e^{ikx},\qquad k^{2}=0.
\end{eqnarray}
{This solution is linearly stable.} It tells us besides $+$ and $\times$ modes propagating at the speed of light, this case allows a breathing mode propagating at the speed of light, corresponding to a massless scalar degree of freedom.

\textbf{\emph{Case 1.4}} $a=0, b\neq0, c\neq0$. Equation (\ref{varphi=}) becomes $(\Box+\frac{c}{b})\varphi=0$ and $\varphi$ has a solution
\begin{eqnarray} 	
	\varphi=\varphi_{0} e^{ikx},\qquad k^{2}=\frac{c}{b}\neq0.
\end{eqnarray}
Since gravitational waves cannot exceed the speed of light, only the parameter selection satisfying $\frac{c}{b}\textless0$ is allowed by physics. {Under this condition, this solution is linearly stable, and} besides $+$ and $\times$ modes propagating at the speed of light, this case also allows a mixture mode of the breathing and longitudinal modes propagating at the speed less than the speed of light, corresponding to a massive scalar degree of freedom.

When $a\neq0$, inspired by the analysis in Refs. \cite{H. Lu1} and \cite{H. Lu2}, we can write Eq. (\ref{varphi=}) in the following form
\begin{eqnarray}
	\label{(Box-a)(Box-b)varphi=0}	
	\left(\Box-{m_{1}}^{2}\right)\left(\Box-{m_{2}}^{2}\right)\varphi=0,
\end{eqnarray}
where
\begin{eqnarray}
	\label{m1^2,m2^2=}	
	\nonumber
	{m_{1}}^{2}=\frac{-b+\sqrt{b^2-4ac}}{2a},
	\\
	{m_{2}}^{2}=\frac{-b-\sqrt{b^2-4ac}}{2a}.
\end{eqnarray}

\textbf{\emph{Case 2.1}} $a\neq0, b^2-4ac\textless0$. In this case ${m_{1}}$ and ${m_{2}}$ are complex, so the solution of $\varphi$ is exponentially divergent except $\varphi=0$. {The exponential growth indicates that the solution is linearly unstable}, so we do not consider it here. The remaining solutions allow only $+$ and $\times$ modes propagating at the speed of light.

%These exponentially divergent solutions are discarded because they contradict the premise that $\varphi$ is a small perturbation. In this case, only $+$ and $\times$ modes propagating at the speed of light are allowed.

\textbf{\emph{Case 2.2}} $a\neq0, b^2-4ac\ge0$. In this case both ${m_{1}}^{2}$ and ${m_{2}}^{2}$ are real. Since superluminal speed is not allowed in physics, it is allowed to be selected only when the parameter ${m_{1}}^{2}, {m_{1}}^{2}\ge0$. This can be further divided into more detailed cases:

\quad\textbf{\emph{Case 2.2.1}} $b=c=0$. In this case ${m_{1}}^{2}={m_{2}}^{2}=0$, so Eq. (\ref{(Box-a)(Box-b)varphi=0}) becomes
\begin{eqnarray}
	\label{Box Box varphi=0}	
	\Box\Box\varphi=0,
\end{eqnarray}
from which we know that $\Box\varphi$ has a solution
   \begin{eqnarray}
		\label{Box varphi=}	
		\Box\varphi=\varphi_{0} e^{ikx},\qquad k^{2}=0.
	\end{eqnarray}
Thus further $\varphi$ has a solution
   \begin{eqnarray}
    \varphi=\left(t+z\right)\frac{\varphi_{0}}{4ik_{3}} e^{ikx}+\varphi'_{0} e^{ik'x},\qquad k^{2}=k'^{2}=0.
\end{eqnarray}
Only the term $\varphi'_{0} e^{ik'x}$ ($\varphi'_{0}$ is constant) does not diverge {and is linearly stable}, so {for the reason described in case 2.1, we only consider } $\varphi_{0}=0$. {Besides} $+$ and $\times$ modes propagating at the speed of light, this case allows a breathing mode propagating at the speed of light, corresponding to a massless scalar degree of freedom.

\quad\textbf{\emph{Case 2.2.2}} $b, c\neq0$ and $b^2-4ac=0$. In this case $m^{2}={m_{1}}^{2}={m_{2}}^{2}=\frac{-b}{2a}\neq0$,  so Eq. (\ref{(Box-a)(Box-b)varphi=0}) becomes
\begin{eqnarray}
	\label{(Box-m)^2 varphi=0}	
	\left(\Box-m^{2}\right)\left(\Box-m^{2}\right)\varphi=0,
\end{eqnarray}
which tells us $\Box\varphi$ has a solution
\begin{eqnarray}
	\label{Box varphi=}	
	\left(\Box-m^{2}\right)\varphi=\varphi_{0} e^{ikx},\qquad k^{2}=-m^{2}.
\end{eqnarray}
Thus the solution of $\varphi$ is
\begin{eqnarray}
	\varphi=\left(k_{3}t-k_{0}z\right)\frac{i\varphi_{0}}{4k_{3}k_{0}} e^{ikx}+\varphi'_{0} e^{ik'x},\qquad k^{2}=k'^{2}=\frac{b}{2a}.
\end{eqnarray}
Similar to case 2.2.1, only the term $\varphi'_{0} e^{ik'x}$ ($\varphi'_{0}$ is constant) does not diverge {and is linearly stable}, so {%for the reason described in case 2.1, we only consider }
	we also consider $\varphi_{0}=0$. Therefore, {besides} $+$ and $\times$ modes propagating at the speed of light, we also have a mixture mode of the breathing and longitudinal modes propagating at the speed less than the speed of light, corresponding to a massive scalar degree of freedom.

\quad\textbf{\emph{Case 2.2.3}} $c=0, b\neq0$. In this case, one of ${m_{1}}^{2}$ and ${m_{2}}^{2}$ is 0 and the other is not. {These solutions are linearly stable.} Therefore, {besides} $+$ and $\times$ modes propagating at the speed of light, there are two scalar degrees of freedom. One is a massless degree of freedom which corresponds to a breathing mode propagating at the speed of light, and the other is a massive degree of freedom which is the mixture of the breathing and longitudinal modes propagating at the speed less than the speed of light.

\quad\textbf{\emph{Case 2.2.4}} $c\neq0, b^2-4ac\neq0$. In this case, ${m_{1}}^{2}\neq{m_{2}}^{2}\neq0$. {These solutions are linearly stable. So} we have $+$ and $\times$ modes propagating at the speed of light, as well as two massive scalar degrees of freedom with different masses. Both modes are the mixture of the breathing and longitudinal modes.

Finally, we summarize the above results in Tab I. {Here we find that in some cases, there are two degrees of freedom in the scalar field, which is different from the metric Horndeski theory. Similar results also appear in Refs. \cite{S. Bahamonde, Flavio Moretti 2}.}

\begin{center}
\begin{table}[htbp]
	\resizebox{\textwidth}{33mm}{
\begin{tabular}{|l|l|c|c|c|c|}
  \hline\hline
  \textbf{Cases} & \qquad\textbf{Conditions} & $+$ mode & $\times$ mode & massless scalar mode & massive scalar mode \\
  \hline
		case 0 & ${\mathop{G}^{0}}~\!\!^{}_{4,\phi}=0.$ & 1 & 1 & 0 & 0 \\
  \hline
		case 1.1 &$\mathop{G}^{0}~\!\!^{}_{4,\phi}\neq0, a=b=c=0.$& - & - & - & - \\
		\hline
		case 1.2 &$\mathop{G}^{0}~\!\!^{}_{4,\phi}\neq0, a=b=0, c\neq0.$& 1 & 1 & 0 & 0 \\
		\hline
		case 1.3 &$\mathop{G}^{0}~\!\!^{}_{4,\phi}\neq0, a=c=0, b\neq0.$& 1 & 1 & 1 & 0 \\
		\hline
		case 1.4 &$\mathop{G}^{0}~\!\!^{}_{4,\phi}\neq0, a=0, b,c\neq0.$& 1 & 1 & 0 & 1 \\
		\hline
		case 2.1 &$\mathop{G}^{0}~\!\!^{}_{4,\phi}\neq0, a\neq0, b^{2}-4ac\textless0.$& 1 & 1 & 0 & 0 \\
		\hline
		case 2.2.1 &$\mathop{G}^{0}~\!\!^{}_{4,\phi}\neq0, a\neq0, b=c=0.$& 1 & 1 & 1 & 0 \\
		\hline
		case 2.2.2 &$\mathop{G}^{0}~\!\!^{}_{4,\phi}\neq0, a\neq0, b,c\neq0,b^{2}-4ac=0 .$& 1 & 1 & 0 & 1 \\
		\hline
		case 2.2.3 &$\mathop{G}^{0}~\!\!^{}_{4,\phi}\neq0, a\neq0, b^{2}-4ac\textgreater0, c=0.$& 1 & 1 & 1 & 1 \\
		\hline
		case 2.2.4 &$\mathop{G}^{0}~\!\!^{}_{4,\phi}\neq0, a\neq0, b^{2}-4ac\textgreater0, c\neq0.$& 1 & 1 & 0 & 2 \\
		\hline
		\hline
\end{tabular}}
     \caption{The number of polarization modes of gravitational waves under various cases. %\textcolor{blue}{(in linear approximation)}.
     	In this table, a massless scalar mode means a breathing mode propagating at the speed of light and a massive scalar mode means the mixture of the breathing and longitudinal modes propagating at the speed less than the speed of light. The number in each mode represents the degrees of freedom of this mode in each case.}
\label{tab:dof_classes}
\end{table}
\end{center}

\section{Conclusion}
\label{sec: 5}
In this paper, {under the background of Minkowski space-time, we study the polarization modes of gravitational waves in Paratini-Horndeski theory under linearized perturbations.} %in the range of a linear analysis.}
%we investigated polarization modes of gravitational waves in Palatini Horndeski theory under the background of the Minkowski space-time.
 We first obtained the field equations satisfied by the Minkowski background solutions of Palatini-Horndeski theory. It was found that when $G_{4}(\phi_{0},0)\neq0$, the quantities $K(\phi_{0},0)$ , $K_{\phi}(\phi_{0},0)$ and the background connection will be zero. Then we analyzed the polarization modes of gravitational waves in this background with the linear equations satisfied by the perturbations.

We found that the polarization modes of gravitational waves in Palatini-Horndeski theory depend on the parameter spaces. The results of various parameter spaces have been summarized in Sec. \ref{sec: 4}. It is worth noting that, different from metric Horndeski theory, Palatini-Horndeski theory allows two scalar degrees of freedom for some parameter spaces, which is caused by the fourth-order equation satisfied by the scalar perturbation $\varphi$. There are two cases for the two scalar degrees of freedom: 1) two massive modes corresponding to different masses. In this case, both modes behave as the mixture of the breathing and longitudinal modes propagating at the speed less than the speed of light; 2) one massive mode and one massless mode. In this case, we can get a breathing mode propagating at the speed of light and a mixture mode of the breathing and longitudinal modes propagating at the speed less than the speed of light. In all cases of the parameter spaces, it is allowed to have $+$ and $\times$ modes propagating at the speed of light but there are no vector modes. Therefore, the polarization modes of gravitational waves predicted by general relativity also appear in Palatini-Horndeski theory.

{We only studied linear perturbations of Minkowski background in Palatini-Horndeski theory. However, it should be pointed out that it is meaningful to investigate the behavior of nonlinear perturbations in this theory under the following cases: case 2.1, case 2.2.1 and case 2.2.2, for which the perturbations are linearly unstable. The nonlinear part of the field equations may inhibit the rapid growth of these linearly unstable solutions, and these solutions may be nonlinear stable.}

%\textcolor{blue}{We only studied the polarization modes of gravitational waves in the range of a linear analysis. However, it should be pointed out that it may be meaningful to investigate the behavior of nonlinear gravitational waves in Palatini Horndeski theory under the following parameters: case 1.1, case 2.1, case 2.2.1 and case 2.2.2.  1) For case 2.1, case 2.2.1 and case 2.2.2, they have divergent gravitational wave solutions under linear approximation. However, the nonlinear part of the equations may inhibit the rapid growth of these solutions, so that these solutions have physical significance. 2) For case 1.1, the equation of scalar field perturbation $\varphi$ is at least quadratic, oo we can not use linear approximation. In this case, there may be nonlinear scalar gravitational waves. The existence of these cases may require us to develop a method to extract the required data from a large number of detected gravitational wave datas when the gravitational waves does not satisfy the linear equations. }

Although the ghost problem has not been fully considered, Ref. \cite{Helpin} studied the existence of Ostrogradsky ghost \cite{M. Ostrogradsky,R. P. Woodard} in Palatini-Horndeski theory under some parameter spaces. As a result, Palatini-Horndeski theory in some parameter spaces can be ghost free, but not all of them are ghost-free. So the requirement that the theory should avoid the Ostrogradsky ghost can be combined with future observation of polarization modes of gravitational waves to further limit the parameter space of Palatini-Horndeski theory.

The polarization modes of gravitational waves in Palatini-Horndeski theory can be divided into quite rich cases by parameter spaces. The appropriate parameter spaces can be expected to be selected in the detection of gravitational wave polarization modes by Lisa, Taiji and TianQin \cite{lisa,taiji,tianqin} in the future.

\section*{Acknowledgments}
We would like to thank Qing-Guo Huang and Yu-Qiang Liu for useful discussion.
This work is supported in part by the National Key Research and Development Program of
China (Grant No. 2020YFC2201503), the National Natural Science Foundation of China
(Grants No. 11875151 and  No. 12047501), the 111 Project (Grant No. B20063) and Lanzhou City's scientific research funding subsidy to Lanzhou University.

\appendix
\section{Second-order perturbation expansion of some quantities}
In this appendix, we list some expressions about the perturbation expansion to second order:

\begin{eqnarray} 	
	\mathop{g^{\mu\nu}}^{(2)}=\eta^{\mu\nu}-h^{\mu\nu}+h^{\mu}_{\rho}h^{\rho\nu},
\end{eqnarray}
\begin{eqnarray} 	
	-\mathop{g}^{(2)}=1+h+\frac{1}{2}h^{2}-\frac{1}{2} h^{\lambda}_{\rho}h^{\rho}_{\lambda}.
\end{eqnarray}
\begin{eqnarray} 	
	\mathop{\sqrt{-g}}^{(2)}=1+\frac{1}{2}h+\frac{1}{8}h^{2}-\frac{1}{4} h^{\lambda}_{\rho}h^{\rho}_{\lambda}.
\end{eqnarray}
\begin{eqnarray} 		\mathop{\mathcal{L}_{2}}^{(2)}=\mathop{K}^{0}+\mathop{K}^{0}~\!\!^{}_{,\phi}\varphi+\mathop{K}^{0}~\!\!^{}_{,X}X+\frac{1}{2}\mathop{K}^{0}~\!\!^{}_{,\phi\phi}\varphi^{2},
\end{eqnarray}
\begin{eqnarray}			\mathop{\mathcal{L}_{3}}^{(2)}&=&-\mathop{G_{3}}^{0}\left(\Box\varphi-\eta^{\mu\nu}\mathop{\Gamma}^{0}~\!\!^{\lambda}_{\nu\mu}\partial_{\lambda}\varphi-\eta^{\mu\nu}\Sigma^{\lambda}_{\nu\mu}\partial_{\lambda}\varphi-h^{\mu\nu}\partial_{\mu}\partial_{\nu}\varphi+h^{\mu\nu}\mathop{\Gamma}^{0}~\!\!^{\lambda}_{\mu\nu}\partial_{\lambda}\varphi\right)	\nonumber \\
	&-&\mathop{G}^{0}~\!\!^{}_{3,\phi} \varphi\left(\Box\varphi-\eta^{\mu\nu}
 \mathop{\Gamma}^{0}~\!\!^{\lambda}_{\nu\mu}\partial_{\lambda}\varphi\right),
\end{eqnarray}
\begin{eqnarray} 		
\mathop{\mathcal{L}_{4}}^{(2)}&=&\mathop{G_{4}}^{0}\,
   \eta^{\mu\nu}
   \left(\partial_{\lambda}\Sigma^{\lambda}_{\mu\nu}
         -\partial_{\nu}\Sigma^{\lambda}_{\mu\lambda}
         +\mathop{\Gamma}^{0}~\!\!^{\lambda}_{\sigma\lambda}
          \mathop{\Gamma}^{0}~\!\!^{\sigma}_{\mu\nu}
         -\mathop{\Gamma}^{0}~\!\!^{\lambda}_{\sigma\nu}
          \mathop{\Gamma}^{0}~\!\!^{\sigma}_{\mu\lambda}\right.
	 \nonumber \\ 	&\quad&+\Sigma^{\lambda}_{\sigma\lambda}\mathop{\Gamma}^{0}~\!\!^{\sigma}_{\mu\nu}+\Sigma^{\sigma}_{\mu\nu}\mathop{\Gamma}^{0}~\!\!^{\lambda}_{\sigma\lambda}-\Sigma^{\lambda}_{\sigma\nu}\mathop{\Gamma}^{0}~\!\!^{\sigma}_{\mu\lambda}-\Sigma^{\sigma}_{\mu\lambda}\mathop{\Gamma}^{0}~\!\!^{\lambda}_{\sigma\nu}
	\nonumber \\ 	&\quad&+\left.\Sigma^{\lambda}_{\sigma\lambda}\Sigma^{\sigma}_{\mu\nu}-\Sigma^{\lambda}_{\sigma\nu}\Sigma^{\sigma}_{\mu\lambda}\right)
 \nonumber \\
&-&\mathop{G_{4}}^{0}h^{\mu\nu}\left(\partial_{\lambda}\Sigma^{\lambda}_{\mu\nu}-\partial_{\nu}\Sigma^{\lambda}_{\mu\lambda}+\mathop{\Gamma}^{0}~\!\!^{\lambda}_{\sigma\lambda}\mathop{\Gamma}^{0}~\!\!^{\sigma}_{\mu\nu}-\mathop{\Gamma}^{0}~\!\!^{\lambda}_{\sigma\nu}\mathop{\Gamma}^{0}~\!\!^{\sigma}_{\mu\lambda}
  \right.
	\nonumber \\ &\quad&+\left. \Sigma^{\lambda}_{\sigma\lambda}\mathop{\Gamma}^{0}~\!\!^{\sigma}_{\mu\nu}+\Sigma^{\sigma}_{\mu\nu}\mathop{\Gamma}^{0}~\!\!^{\lambda}_{\sigma\lambda}-\Sigma^{\lambda}_{\sigma\nu}\mathop{\Gamma}^{0}~\!\!^{\sigma}_{\mu\lambda}-\Sigma^{\sigma}_{\mu\lambda}\mathop{\Gamma}^{0}~\!\!^{\lambda}_{\sigma\nu}\right)
	\nonumber \\
    &+&\mathop{G_{4}}^{0}h^{\mu}_{\rho}h^{\rho\nu}\left(\mathop{\Gamma}^{0}~\!\!^{\lambda}_{\sigma\lambda}\mathop{\Gamma}^{0}~\!\!^{\sigma}_{\mu\nu}-\mathop{\Gamma}^{0}~\!\!^{\lambda}_{\sigma\nu}\mathop{\Gamma}^{0}~\!\!^{\sigma}_{\mu\lambda}\right)
    \nonumber\\
    &+&\mathop{G}^{0}~\!\!^{}_{4,\phi}\varphi\eta^{\mu\nu}\left(\partial_{\lambda}\Sigma^{\lambda}_{\mu\nu}-\partial_{\nu}\Sigma^{\lambda}_{\mu\lambda}+\mathop{\Gamma}^{0}~\!\!^{\lambda}_{\sigma\lambda}\mathop{\Gamma}^{0}~\!\!^{\sigma}_{\mu\nu}-\mathop{\Gamma}^{0}~\!\!^{\lambda}_{\sigma\nu}\mathop{\Gamma}^{0}~\!\!^{\sigma}_{\mu\lambda}\right.
    \nonumber\\
    &\quad&+\left. \Sigma^{\lambda}_{\sigma\lambda}\mathop{\Gamma}^{0}~\!\!^{\sigma}_{\mu\nu}+\Sigma^{\sigma}_{\mu\nu}\mathop{\Gamma}^{0}~\!\!^{\lambda}_{\sigma\lambda}-\Sigma^{\lambda}_{\sigma\nu}\mathop{\Gamma}^{0}~\!\!^{\sigma}_{\mu\lambda}-\Sigma^{\sigma}_{\mu\lambda}\mathop{\Gamma}^{0}~\!\!^{\lambda}_{\sigma\nu}\right)
    \nonumber\\
    &-&\mathop{G}^{0}~\!\!^{}_{4,\phi}\varphi h^{\mu\nu}\left(\mathop{\Gamma}^{0}~\!\!^{\lambda}_{\sigma\lambda}\mathop{\Gamma}^{0}~\!\!^{\sigma}_{\mu\nu}-\mathop{\Gamma}^{0}~\!\!^{\lambda}_{\sigma\nu}\mathop{\Gamma}^{0}~\!\!^{\sigma}_{\mu\lambda}\right)
    \nonumber\\
    &+&\mathop{G}^{0}~\!\!^{}_{4,X}X\eta^{\mu\nu}\left(\mathop{\Gamma}^{0}~\!\!^{\lambda}_{\sigma\lambda}\mathop{\Gamma}^{0}~\!\!^{\sigma}_{\mu\nu}-\mathop{\Gamma}^{0}~\!\!^{\lambda}_{\sigma\nu}\mathop{\Gamma}^{0}~\!\!^{\sigma}_{\mu\lambda}\right)
    \nonumber\\ &+&\frac{1}{2}\mathop{G}^{0}~\!\!^{}_{4,\phi\phi}\varphi^{2}\eta^{\mu\nu}\left(\mathop{\Gamma}^{0}~\!\!^{\lambda}_{\sigma\lambda}\mathop{\Gamma}^{0}~\!\!^{\sigma}_{\mu\nu}-\mathop{\Gamma}^{0}~\!\!^{\lambda}_{\sigma\nu}\mathop{\Gamma}^{0}~\!\!^{\sigma}_{\mu\lambda}\right)
    \nonumber\\
    &+&\mathop{G}^{0}~\!\!^{}_{4,X}\left[\left(\Box\varphi\right)^2-2\eta^{\mu\nu}\mathop{\Gamma}^{0}~\!\!^{\lambda}_{\nu\mu}\partial_{\lambda}\varphi\Box\varphi+\eta^{\mu\nu}\eta^{\alpha\beta}\mathop{\Gamma}^{0}~\!\!^{\lambda}_{\nu\mu}\mathop{\Gamma}^{0}~\!\!^{\rho}_{\beta\alpha}\partial_{\lambda}\varphi\partial_{\rho}\varphi\right.
    \nonumber\\
    &\quad&-\partial_{\mu}\partial_{\nu}\varphi\partial^{\mu}\partial^{\nu}\varphi-\mathop{\Gamma}^{0}~\!\!^{\nu}_{\sigma\lambda}\partial^{\lambda}\partial_{\nu}\varphi\partial^{\sigma}\varphi+\mathop{\Gamma}^{0}~\!\!^{\sigma}_{\nu\mu}\partial_{\sigma}\varphi\partial^{\mu}\partial^{\nu}\varphi
    \nonumber\\
    &\quad&+\left. \eta^{\mu\lambda}\mathop{\Gamma}^{0}~\!\!^{\sigma}_{\nu\mu}\mathop{\Gamma}^{0}~\!\!^{\nu}_{\gamma\lambda}\partial_{\sigma}\varphi\partial^{\gamma}\varphi\right],
\end{eqnarray}
\begin{eqnarray}
    \nonumber	\mathop{\mathcal{L}_{5}}^{(2)}
 &=& \mathop{G_{5}}^{0}
     \left[\mathop{\Gamma}^{0}~\!\!^{\lambda}_{\sigma\lambda}\mathop{\Gamma}^{0}~\!\!^{\sigma}_{\mu\nu}
            -\mathop{\Gamma}^{0}~\!\!^{\lambda}_{\sigma\nu}\mathop{\Gamma}^{0}~\!\!^{\sigma}_{\mu\lambda}
            -\frac{1}{2}\eta_{\mu\nu}\eta^{\alpha\beta}
             \left(\mathop{\Gamma}^{0}~\!\!^{\lambda}_{\sigma\lambda}
                    \mathop{\Gamma}^{0}~\!\!^{\sigma}_{\alpha\beta}
                   -\mathop{\Gamma}^{0}~\!\!^{\lambda}_{\sigma\beta}
                    \mathop{\Gamma}^{0}~\!\!^{\sigma}_{\alpha\lambda}
             \right)
     \right]
	\\ \nonumber
  &\quad&\times
      \left(\partial^{\mu}\partial^{\nu}\varphi+\eta^{\mu\kappa}
            \mathop{\Gamma}^{0}~\!\!^{\nu}_{\delta\kappa}\partial^{\delta}\varphi
            +\eta^{\mu\kappa}\Sigma^{\nu}_{\delta\kappa}\partial^{\delta}\varphi
            -h^{\mu\kappa}\partial_{\kappa}\partial^{\nu}\varphi
            -h^{\mu\kappa}\mathop{\Gamma}^{0}~\!\!^{\nu}_{\delta\kappa}\partial^{\delta}\varphi\right.
	\\ \nonumber	
 &\quad&
    \left.-\partial^{\mu}  h^{\nu\delta}  \partial_{\delta}\varphi
          -h^{\nu\delta}\partial^{\mu}\partial_{\delta}\varphi
          -\eta^{\mu\kappa}\mathop{\Gamma}^{0}~\!\!^{\nu}_{\delta\kappa}
            h^{\delta\gamma}\partial_{\gamma}\varphi\right)
	\\ \nonumber	
  &+&\mathop{G_{5}}^{0}
    \left[\partial_{\lambda}\Sigma^{\lambda}_{\mu\nu}-\partial_{\nu}\Sigma^{\lambda}_{\mu\lambda}
          +\Sigma^{\lambda}_{\sigma\lambda}\mathop{\Gamma}^{0}~\!\!^{\sigma}_{\mu\nu}
          +\Sigma^{\sigma}_{\mu\nu}\mathop{\Gamma}^{0}~\!\!^{\lambda}_{\sigma\lambda}
          -\Sigma^{\lambda}_{\sigma\nu}\mathop{\Gamma}^{0}~\!\!^{\sigma}_{\mu\lambda}
          -\Sigma^{\sigma}_{\mu\lambda}\mathop{\Gamma}^{0}~\!\!^{\lambda}_{\sigma\nu}
    \right.  	\\ \nonumber	
  &\quad&-\frac{1}{2}\eta_{\mu\nu}\eta^{\alpha\beta}
          \left(\partial_{\lambda}\Sigma^{\lambda}_{\alpha\beta}
                -\partial_{\beta}\Sigma^{\lambda}_{\alpha\lambda}
                +\Sigma^{\lambda}_{\sigma\lambda}\mathop{\Gamma}^{0}~\!\!^{\sigma}_{\alpha\beta}
                +\Sigma^{\sigma}_{\alpha\beta}\mathop{\Gamma}^{0}~\!\!^{\lambda}_{\sigma\lambda}
                -\Sigma^{\lambda}_{\sigma\beta}\mathop{\Gamma}^{0}~\!\!^{\sigma}_{\alpha\lambda}
                -\Sigma^{\sigma}_{\alpha\lambda}\mathop{\Gamma}^{0}~\!\!^{\lambda}_{\sigma\beta}
          \right) 	\\ \nonumber	
 &\quad&+ 
   \left. \frac{1}{2}\eta_{\mu\nu}h^{\alpha\beta}
          \left(\mathop{\Gamma}^{0}~\!\!^{\lambda}_{\sigma\lambda}\mathop{\Gamma}^{0}~\!\!^{\sigma}_{\alpha\beta}
              -\mathop{\Gamma}^{0}~\!\!^{\lambda}_{\sigma\beta}\mathop{\Gamma}^{0}~\!\!^{\sigma}_{\alpha\lambda}
          \right)
          -\frac{1}{2}\eta^{\alpha\beta}h_{\mu\nu}
            \left(\mathop{\Gamma}^{0}~\!\!^{\lambda}_{\sigma\lambda}\mathop{\Gamma}^{0}~\!\!^{\sigma}_{\alpha\beta}
                  -\mathop{\Gamma}^{0}~\!\!^{\lambda}_{\sigma\beta}\mathop{\Gamma}^{0}~\!\!^{\sigma}_{\alpha\lambda}
            \right)
   \right]
	\\ \nonumber	
 &\quad&\times
  \left(\partial^{\mu}\partial^{\nu}\varphi
        +\eta^{\mu\kappa}\mathop{\Gamma}^{0}~\!\!^{\nu}_{\delta\kappa}\partial^{\delta}\varphi
  \right)
	\\ \nonumber	
&+&\mathop{G}^{0}~\!\!^{}_{5,\phi}\varphi
  \left[\mathop{\Gamma}^{0}~\!\!^{\lambda}_{\sigma\lambda}\mathop{\Gamma}^{0}~\!\!^{\sigma}_{\mu\nu}
        -\mathop{\Gamma}^{0}~\!\!^{\lambda}_{\sigma\nu}\mathop{\Gamma}^{0}~\!\!^{\sigma}_{\mu\lambda}
        -\frac{1}{2}\eta_{\mu\nu}\eta^{\alpha\beta}
         \left(\mathop{\Gamma}^{0}~\!\!^{\lambda}_{\sigma\lambda}
               \mathop{\Gamma}^{0}~\!\!^{\sigma}_{\alpha\beta}
              -\mathop{\Gamma}^{0}~\!\!^{\lambda}_{\sigma\beta}
               \mathop{\Gamma}^{0}~\!\!^{\sigma}_{\alpha\lambda}
         \right)
  \right]
	\\	&\quad&\times
  \left(\partial^{\mu}\partial^{\nu}\varphi+\eta^{\mu\kappa}
        \mathop{\Gamma}^{0}~\!\!^{\nu}_{\delta\kappa}\partial^{\delta}\varphi
  \right).
\end{eqnarray}


\begin{thebibliography}{99}
\bibitem{Abbott1}
B. P. Abbott et al. (LIGO Scientific Collaboration and Virgo Collaboration),
``Observation of Gravitational Waves from a Binary Black Hole Merger",
\textit{Phys. Rev. Lett}. \textbf{116}, 061102 (2016).

\bibitem{Abbott2}
B. P. Abbott et al. (LIGO Scientific Collaboration and Virgo Collaboration),
``GW151226: Observation of Gravitational Waves from a 22-Solar-Mass Binary
Black Hole Coalescence",
\textit{Phys. Rev. Lett}. \textbf{116}, 241103 (2016).

\bibitem{Abbott3}
B. P. Abbott et al. (LIGO Scientific Collaboration and Virgo Collaboration),
``GW170104: Observation of a 50-Solar-Mass Binary Black Hole Coalescence at Redshift 0.2",
\textit{Phys. Rev. Lett}. \textbf{118}, 221101 (2017).

\bibitem{Abbott4}
B. P. Abbott et al. (LIGO Scientific Collaboration and Virgo Collaboration),
``GW170814: A Three-Detector Observation of Gravitational Waves from a Binary Black Hole Coalescence",
 \textit{Phys. Rev. Lett}. \textbf{119}, 141101 (2017).

\bibitem{Abbott5}
B. P. Abbott et al. (LIGO Scientific Collaboration and Virgo Collaboration),
``GW170817: Observation of Gravitational Waves from a Binary Neutron Star Inspiral",
 \textit{Phys. Rev. Lett}. \textbf{119}, 161101 (2017).





\bibitem{MTW}
C. W. Misner, K. S. Thorne, and J. A. Wheeler,
``Gravitation",
(\textit{Princeton University Press},  1973)






\bibitem{Eardley}
D. M. Eardley, D. L. Lee and A. P. Lightman,
``Gravitational-wave observations as a tool for testing relativistic gravitye",
\textit{Phys. Rev. D} \textbf{8}, 3308 (1973).



\bibitem{Hiroki Takeda}
H. Takeda, S. Morisaki, and A. Nishizawa,
``Pure polarization test of GW170814 and GW170817 using waveforms consistent with modified theories of gravity",
\textit{Phys. Rev. D}. \textbf{103}, 064037 (2021).


\bibitem{B. P . Abbottet al.2}
{B. P. Abbott et al.} (LIGO Scientific Collaboration and Virgo Collaboration),
``Search for Tensor, Vector, and Scalar Polarizations in the Stochastic Gravitational-Wave Background",
\textit{Phys. Rev. Lett}. \textbf{120}, 201102 (2018).

\bibitem{B. P . Abbottet al.3}
{B. P. Abbott et al.} (LIGO Scientific Collaboration and Virgo Collaboration),
``Search for the isotropic stochastic background using data from Advanced LIGO’s second observing run",
\textit{Phys. Rev. D}. \textbf{100}, 061101(R) (2019).

\bibitem{B. P . Abbottet al.4}
{B. P. Abbott et al.} (LIGO Scientific Collaboration and Virgo Collaboration),
``Tests of general relativity with binary black holes from the second LIGO-Virgo gravitational-wave transient catalog",
\textit{Phys. Rev. D}. \textbf{103}, 122002 (2021).



\bibitem{Atsushi Nishizawa}
A. Nishizawa, A. Taruya, K. Hayama, S. Kawamura, and M. A. Sakagami,
``Probing nontensorial polarizations of stochastic gravitational-wave backgrounds with ground-based laser interferometers",
\textit{Phys. Rev. D} \textbf{79}, 082002 (2009).

\bibitem{Kazuhiro Hayama}
K. Hayama and A. Nishizawa,
``Model-independent test of gravity with a network of ground-based gravitational-wave detectors",
\textit{Phys. Rev. D} \textbf{87}, 062003 (2013).

\bibitem{Maximiliano Isi1}
M. Isi, A. J. Weinstein, C. Mead and M. Pitkin,
``Detecting beyond-Einstein polarizations of continuous gravitational waves",
\textit{Phys. Rev. D} \textbf{91}, 082002 (2015).

\bibitem{Maximiliano Isi2}
M. Isi, M. Pitkin and A. J. Weinstein,
``Probing dynamical gravity with the polarization of continuous gravitational waves",
\textit{Phys. Rev. D} \textbf{96}, 042001 (2017).






\bibitem{K. Chatziioannou}
K. Chatziioannou, N. Yunes and N. Cornish,
``Model-independent test of general relativity: An extended post-Einsteinian framework with complete polarization content",
\textit{Phys. Rev. D} \textbf{86}, 022004 (2012) [Erratum: Phys. Rev. D 95, 129901(E) (2017)]

\bibitem{H. Takeda}
H. Takeda, A. Nishizawa, Y. Michimura, K. Nagano, K. Komori, M. Ando and K. Hayama,
``Polarization test of gravitational waves from compact binary coalescences",
\textit{Phys. Rev. D} \textbf{98}, 022008 (2018).





\bibitem{Yuki Hagihara}
Y. Hagihara, N. Era, D. Iikawa, and H. Asada,
``Probing gravitational wave polarizations with Advanced LIGO, Advanced Virgo, and KAGRA",
\textit{Phys. Rev. D} \textbf{98}, 064035 (2018).

\bibitem{P. T. H. Pang}
P. T. H. Pang, R. K. L. Lo, I. C. F. Wong, T. G. F. Li, and C. VanDenBroeck,
``Generic searches for alternative gravitational wave polarizations with networks of interferometric detectors",
\textit{Phys. Rev. D} \textbf{101}, 104055 (2020).

\bibitem{B. P . Abbottet al.1}
{B. P. Abbott et al.} (LIGO Scientific Collaboration and Virgo Collaboration),
``First Search for Nontensorial Gravitational Waves from Known Pulsars",
 \textit{Phys. Rev. Lett}. \textbf{120}, 031104 (2018).



%\bibitem{Hiroki Takeda}
%H. Takeda, S. Morisaki and A. Nishizawa,
%``Pure polarization test of GW170814 and GW170817 using waveforms consistent with modified theories of gravity",
%\textit{Phys. Rev. D}. \textbf{103}, 064037 (2021).











%\bibitem{H. Takeda}
%H. Takeda, A. Nishizawa, Y. Michimura, K. Nagano, K. Komori, M. Ando and K. Hayama,
%``Polarization test of gravitational waves from compact binary coalescences",
%\textit{Phys. Rev. D} \textbf{98}, 022008 (2018).

\bibitem{f(R}
D. Liang, Y. Gong, S. Hou and Y. Liu,
``Polarizations of gravitational waves in $f(R)$ gravity",
\textit{Phys. Rev. D} \textbf{95}, 104034 (2017).

\bibitem{Horndeski}
S. Hou, Y. Gong and Y. Liu,
``Polarizations of gravitational waves in Horndeski theory",
\textit{Eur. Phys. J. C} \textbf{78}, 378 (2018).

\bibitem{TeVeS}
Y. Gong, S. Hou, D. Liang and E. Papantonopoulos,
``Gravitational waves in Einstein-ather and generalized TeVeS theory after GW170817",
\textit{Phys. Rev. D} \textbf{97}, 084040  (2018).

\bibitem{Horava}
Y. Gong, S. Hou, E. Papantonopoulos and D. Tzortzis,
``Gravitational waves and the polarizations in Horava gravity after GW170817",
 \textit{Phys. Rev. D} \textbf{98}, 104017  (2018).

\bibitem{STVG}
Y. Liu, W. Qian, Y. Gong and B. Wang,
``Gravitational waves in scalar-tensor-vector gravity theory",
 \textit{Universive} \textbf{7}, 9 (2021).

\bibitem{fT}
K. Bamba, S. Capozziello, M. D. Laurentis, S. Nojiri and D. Saez-Gomez,
``No further gravitational wave modes in $f(T)$ gravity",
 \textit{Phys. Lett. B} \textbf{727}, 194 (2013).

\bibitem{dCS and EdGB}
P. Wagle, A. Saffer and N. Yunes,
``Polarization modes of gravitational waves in quadratic gravity",
\textit{Phys. Rev. D} \textbf{100}, 124007 (2019).

\bibitem{N-P}
E. Newman and R. Penrose,
``An approach to gravitational radiation by a method of spin coefficients",
 \textit{J. Math. Phys}. \textbf{3}, 566-578 (1962).

%\bibitem{lisa}
%E. Belgacem et al. (LISA Cosmology Working Group Collaboration),
%``Testing modified gravity at cosmological distances with LISA standard sirens",
% \textit{J. Cosmol. Astropart. Phys}. \textbf{7}, 024 (2019).

\bibitem{lisa}
 P. Amaro-Seoane, H. Audley, S. Babak and et al. (LISA Team),
``Laser interferometer space antenna",
eprint arXiv:1702.00786.

\bibitem{taiji}
Z. Luo, Y. Wang, Y. Wu, W. Hu and G. Jin,
``The Taiji program: A concise overview",
\textit{Prog. Theor. Exp. Phys}. \textbf{2021}, 05A108 (2021).

\bibitem{Lisa-taiji}
G. Wang and W. B. Han,
``Observing gravitational wave polarizations with the LISA-TAIJI network",
\textit{Phys. Rev. D}. \textbf{103}, 064021 (2021).

\bibitem{tianqin}
J. Luo, L. S. Chen, H. Z. Duan et al.,
``TianQin: A space-borne gravitational wave detector",
\textit{Classical. Quant. Grav}. \textbf{33}, 035010 (2016).

\bibitem{Z. Chen}
{Z. Chen, C. Yuan and Q. Huang,
	``Non-tensorial gravitational wave background in NANOGrav 12.5-year data set",
	\textit{Sci. China Phys. Mech. Astron.} \textbf{64}, 120412 (2021).}

\bibitem{Horndeski2}
 G. W. Horndeski,
 ``Second-order scalar-tensor field equations in a four-dimensional space",
 \textit{Int. J Theor. Phys}. \textbf{10}, 363 (1974).

\bibitem{Horndeski3}
  T. Kobayashi, M. Yamaguchi and J. Yokoyama,
  ``Generalized G-inflation: Inflation with the most general second-order field equations",
   \textit{Prog. Theor. Phys}. \textbf{126}, 511  (2011).

\bibitem{Fabio Moretti 1}
{F. Moretti, F. Bombacigno and G. Montani,
``Gravitational Landau Damping for massive scalar modes",
\textit{Eur. Phys. J. C}. \textbf{80}, 1203  (2020).}

\bibitem{Harvey S. Reall}
{H. S. Reall,
``Causality in gravitational theories with second order equations of
motion",
\textit{Phys. Rev. D}. \textbf{103}, 084027 (2021).}

\bibitem{Helpin}
  T. Helpin and M. S. Volkov,
  ``Varying the Horndeski Lagrangian within the Palatini approach",
   \textit{J. Cosmol. Astropart. Phys.} \textbf{01}, 044 (2020).

\bibitem{Helpin2}
  T. Helpin and M. S. Volkov,
  ``A metric-affine version of the Horndeski theory",
   \textit{Int J Mod. Phys. A} \textbf{35}, 2040010 (2020).

\bibitem{Shimada}
 K. Shimada, K. Aoki and K. I. Maeda,
 ``Metric-affine gravity and inflation",
  \textit{Phys. Rev. D} \textbf{99}, 104020 (2019).

\bibitem{B. P. Abbott000}
B. P. Abbottet et al. (LIGO Scientific Collaboration and Virgo Collaboration),
``Gravitational waves and gamma-rays from a binary neutron star merger: GW170817 and GRB 170817A",
\textit{Astrophys. J.} \textbf{848}, L13 (2017).


\bibitem{B. P. Abbott111}
B. P. Abbottet et al. (LIGO Scientific Collaboration and Virgo Collaboration),
``Tests of General Relativity with GW170817",
\textit{Phys. Rev. Lett.} \textbf{123}, 011102 (2019).

\bibitem{P. Creminelli and F. Vernizzi}
{P. Creminelli and F. Vernizzi,
	``Dark Energy after GW170817 and GRB170817A",
	\textit{Phys. Rev. Lett.} \textbf{119}, 251302 (2017).}



\bibitem{C. D. Kreisch}
{C. D. Kreisch and E. Komatsu,
	``Cosmological constraints on Horndeski gravity in light of GW170817",
	\textit{J. Cosmol. Astropart. Phys.} \textbf{12}, 030 (2018).}

\bibitem{Y. Gong 00}
{Y. Gong, E. Papantonopoulos, and Z. Yi,
	``Constraints on scalar-tensor theory of gravity by the recent observational results on gravitational waves",
	\textit{Eur. Phys. J. C} \textbf{78}, 738 (2018).}

\bibitem{S. Bahamonde 1}
{S. Bahamonde, K. F. Dialektopoulos and J. L. Said,
	``Can {Horndeski} theory be recast using teleparallel gravity?",
	\textit{Phys. Rev. D} \textbf{100}, 064018 (2019).}

\bibitem{S. Bahamonde 2}
{S. Bahamonde, K. F. Dialektopoulos, V. Gakis and J. L. Said,
	``Reviving Horndeski theory using teleparallel gravity after GW170817",
	\textit{Phys. Rev. D} \textbf{101}, 084060 (2020).}

\bibitem{F. Bombacigno}
F. Bombacigno and G. Montani,
``Implications of the Holst term in a $f(R)$ theory with torsion",
 \textit{Phys. Rev. D} \textbf{99}, 064016 (2019).

\bibitem{J.Lu}
J. Lu, J. Li, H. Guo, Z. Zhuang and X. Zhao,
``Linearized physics and gravitational-waves polarizations in the Palatini formalism of GBD theory",
 \textit{Phys. Lett. B} \textbf{811}, 135985 (2020).

\bibitem{S. Bahamonde}
S. Bahamonde, and M. Hohmann, M. Caruana, K. F. Dialektopoulos, V. Gakis, J. LeviSaid, E. N. Saridakis and J. Sultana,
``Gravitational-wave propagation and polarizations in the teleparallel analog of Horndeski gravity",
 \textit{Phys. Rev. D} \textbf{104}, 084082 (2021).

\bibitem{Einstein}
 A. Einstein,
 ``Einheitliche Feldtheorie von Gravitation und Elektrizit\"{a}t",
 \textit{Sitzungsber. Preuss. Akad. Wiss}. \textbf{22}, 414 (1925).

\bibitem{Palatini1}
  A. Palatini,
  ``Deduzione invariantiva delle equazioni gravitazionali dal principio di Hamilton",
   \textit{Rend. Circ. Mat. Palermo} \textbf{43}, 203 (1919).

\bibitem{Palatini2}
 M. Ferraris, M. Francaviglia and C. Reina,
 ``Variational formulation of general relativity from 1915 to 1925 `Palatini's method' discovered by Einstein in 1925",
 \textit{Gen. Relat. Gravit}. \textbf{14}, 243 (1982).

\bibitem{TH. Hyun}
 Y. H. Hyun, Y. Kim and S. Lee,
 ``Exact amplitudes of six polarization modes for gravitational waves",
  \textit{Phys. Rev. D} \textbf{99}, 124002 (2019).

\bibitem{Fabio Moretti 3}
{F. Moretti, F. Bombacigno and G. Montani,
``Gauge invariant formulation of metric $f(R)$ gravity for gravitational waves",
\textit{Phys. Rev. D} \textbf{100}, 084014 (2019).  }

\bibitem{H. Lu1}
H. Lu and C. N. Pope,
``Critical Gravity in Four Dimensions",
 \textit{Phys. Rev. Lett}. \textbf{106}, 181302 (2011).

\bibitem{H. Lu2}
S. Deser, H. Liu, H. Lu, C. N. Pope, T. C.  Sisman and B. Tekin,
``Critical points of D-Dimensional extended gravities",
 \textit{Phys. Rev. D} \textbf{83}, 061502(R).

\bibitem{Flavio Moretti 2}
{F. Bombacigno, F. Moretti and G. Montani,
``Scalar modes in extended hybrid metric-Palatini gravity: Weak field phenomenology",
\textit{Phys. Rev. D} \textbf{100}, 124036 (2019).}

\bibitem{M. Ostrogradsky}
M. Ostrogradsky,
``Memoires sur les equations differentielle relatives au probleme des
isoperimetres",
 \textit{Mem. Ac. St. Petersbourg} \textbf{VI 4}, 385 (1850).

\bibitem{R. P. Woodard}
R. P. Woodard,
``Avoiding dark energy with $1/R$ modifications
of gravity",
 \textit{Lect. Notes. Phys}. \textbf{720}, 403 (2007).
\end{thebibliography}
\end{document}